\documentclass[sigconf]{acmart}%

\AtBeginDocument{%
  \providecommand\BibTeX{{%
    \normalfont B\kern-0.5em{\scshape i\kern-0.25em b}\kern-0.8em\TeX}}}

\usepackage{bm}
\usepackage{multirow,algorithmic,subcaption}
\usepackage[algoruled,linesnumbered]{algorithm2e}

\begin{document}

\title{GAGE: Geometry Preserving Attributed Graph Embeddings}

\author{Charilaos I. Kanatsoulis}
\affiliation{%
  \institution{University of Pennsylvania}
  \city{Philadelphia}
  \state{Pennsylvania}
  \country{USA}}
\email{kanac@seas.upenn.edu}

\author{Nicholas D. Sidiropoulos}
\affiliation{%
  \institution{University of Virginia}
  \city{Charlottesville}
  \state{Virginia}
  \country{USA}}
\email{nikos@virginia.edu}

\newtheorem{Theorem}{Theorem}
\newtheorem{Result}{Result}
\begin{abstract}
Node embedding is the task of extracting concise and informative representations of certain entities that are connected in a network. Various real-world networks include information about both node connectivity and certain node attributes, in the form of features or time-series data. Modern representation learning techniques employ both the connectivity and attribute information of the nodes to produce embeddings in an unsupervised manner. In this context, deriving embeddings that preserve the geometry of the network and the attribute vectors would be highly desirable, as they would reflect both the topological neighborhood structure and proximity in feature space. While this is fairly straightforward to maintain when only observing the connectivity or attribute information of the network, preserving the geometry of both types of information is challenging. A novel tensor factorization approach for node embedding in attributed networks is proposed in this paper, that preserves the distances of both the connections and the attributes. Furthermore, an effective and lightweight algorithm is developed to tackle the learning task and judicious experiments with multiple state-of-the-art baselines suggest that the proposed algorithm offers significant performance improvements in downstream tasks.  
\end{abstract}


\maketitle

\section{Introduction}
Network science studies the behavior of entities, belonging to one or more communities, via observing their mutual interactions \cite{barabasi2016network}. Networks and network science have attracted considerable attention in science and engineering, since they offer an elegant abstraction of various physical, social, and engineered systems -- and effective tools to analyze them \cite{easley2010networks,newman2018networks}. Networks are nowadays ubiquitous in a plethora of science and engineering disciplines, including social, communication, and biological networks, to name a few. 

Networks are usually represented by graphs, which are informative abstractions and model the interactions in the system. In particular, graph representations encode the connectivity information of different entities (nodes) through a set of edges. The connectivity information in a network is important and describes each node in the network with respect to the rest of the nodes. In real world networks, the entities are not only defined by their connectivity with other entities, but can also be described by a set of measurements or attributes, which offer a node characterisation at an individual level, and are usually very informative. Although graphs offer an elegant representation of the network entities, individual representations of the entities is also required that is not necessarily described by relations with respect to subsets of the community. Furthermore, when attributes are available for each node, which is often the case in practice, it is essential to combine both connectivity and attribute information in a single, universal representation, that encapsulates as much information as possible. Moreover, a variety of interesting networks involve millions of nodes, which makes graph representation of nodes highly impractical for certain tasks.

The aforementioned challenges underscore the need for concise and informative representation of network nodes that is conducive for exploratory analysis, as well as downstream applications. This has motivated a considerable body of research on embedding graph nodes in a low-dimensional vector space, using graph and attribute information in an unsupervised manner. The task is also known as unsupervised node or graph representation learning. The objective of unsupervised node embedding is twofold. On the one hand, the embeddings should capture the maximum amount of knowledge present in the graph and attributes so that information loss is avoided. Towards this end, a key to successful node embeddings is to be able to preserve the geometry of the network, defined by proximity in both the connectivity and the attributes of the nodes. On the other hand the embedding should be able to boost the performance of various downstream network tasks, such as node classification, link prediction, and community detection, to name a few. Concise node representations produced by embedding algorithms can significantly benefit feature-based tools such as logistic regression, support vector machines, and even neural networks -- especially when only limited training data are accessible. 

A plethora of methods have been proposed to perform node embedding. Early works approached the node representation learning task using only the connectivity information of the network. A number of them focused on properly defining a similarity measure on the connectivity information and performing matrix factorization on it \cite{ahmed2013distributed,yang2015network,ou2016asymmetric,qiu2018network,cao2015grarep,shaw2009structure,tang2015line,tsitsulin2018verse,berberidis2019node}. The work in \cite{kanatsoulis2021tex} performs coupled tensor-matrix factorization to learn representations in the context of knowledge graphs. Random walks have also been successfully employed to generate node embeddings, e.g., \cite{perozzi2014deepwalk,grover2016node2vec}. More recently, the focus of research has shifted towards generating embeddings for attributed networks. The work in \cite{yang2015network} generalizes deepwalk \cite{perozzi2014deepwalk} to the case where attributes are available, while \cite{huang2017label} performs label-informed attributed node representation learning in a semi-supervised setting. Neural networks for network science tasks have also gained significant attention lately. In particular, graph convolutional neural networks and graph auto-encoders are very popular for attributed node embedding \cite{cao2016deep,wang2016structural,kipf2016semi,kipf2016variational,velickovic2019deep,gao2018deep,cui2020adaptive}. Works have also been proposed to perform inductive embedding, e.g., \cite{hamilton2017inductive,velickovic2019deep} where a graph convolutional network is trained with multiple graphs. Finally, the work in \cite{al2018t} employs a tensor decomposition model and jointly factors the conventional adjacency along with a $k$-nearest neighbor matrix of the attributes.

Our work is motivated by the following question: {\em Can we produce node embeddings such that we provably preserve the geometry of 1) the distances associated with the connectivity information of the network, and 2) the distances associated with the attributed information of the network, in an unsupervised manner?} This is a well motivated problem, since maintaining the network geometry is a fundamental objective of representation learning, and, as we will show in this paper, doing so significantly improves the performance of several downstream tasks. The problem can be informally stated as follows:
	\begin{itemize}
		\itemsep0em
		\item {\bf Given:} the connectivity and attribute information of network nodes. 
		\item {\bf Produce:} Low dimensional node representations that preserve both the connectivity and attribute geometry.
	\end{itemize}
We propose {\sf Geometry-preserving Attributed Graph Embedding (GAGE)} -- a principled approach to extract node embeddings in an unsupervised fashion. GAGE enjoys several favorable properties.
\begin{itemize}
    \item By design, the produced embeddings preserve node geometry, as inferred from both the node adjacency matrix and the node attributes.
    \item The node embeddings are unique and thus permutation invariant, meaning that any reordering of the nodes in the adjacency representation yield the same embeddings.
    \item The approach is applicable to both undirected and directed networks. 
    \item The proposed approach is flexible and does not require connectivity and attribute information for every node; embeddings can be produced for nodes with partially/completely missing connectivity {\em or} attribute information (but not both).  
    \item The proposed algorithm is lightweight and scalable -- it can efficiently handle large networks.
\end{itemize}


To assess the performance of \texttt{GAGE}, we used three real attributed network benchmarks. Experimental results show that \texttt{GAGE} shows great promise in both tasks, markedly outperforming the baselines.
The contributions of our work can be summarized as follows:
\begin{itemize}
	\item {\bf Novel problem formulation}: Previous work in this area hasn't formalized the intuitive requirement that the embedding should be capable of (approximately) reproducing the distances associated with the connectivity and attribute information. 
	\item {\bf Analysis:} We show that by leveraging the favorable properties of tensor factorization and multi dimensional scaling the proposed embedding can  (approximately) reproduce both the connectivity and attribute distances.
	\item {\bf Algorithm:} We propose a novel tensor factorization algorithm to perform unsupervised embedding task. The algorithm exploits the special structure of the tensor, is fast and scalable for big networks.
	\item {\bf Experimental verification:} The proposed approach is assessed under node classification and link prediction settings and exhibits very promising results in both tasks.
\end{itemize}

\noindent{\bf Reproducibility:} The datasets we use are all publicly available; Code and data can be found in the following link \footnote{https://github.com/marhar19/GAGE-Geometry-Preserving-Attributed-Graph-Embeddings}.


\noindent \textbf{Notation:} Our notation is summarized in Table \ref{tab:TableOfNotationForMyResearch}.
\vspace{-0.1cm}
\begin{table}[ht]\caption{Overview of notation.}
\vspace{-0.2cm}
	\centering 
	{\small\begin{tabular}{r c p{6cm} }
		\toprule
		$\mathcal{V}$ & $\triangleq$ & Set of nodes\\
		$\mathcal{E}$ & $\triangleq$ & Set of edges\\		
		$\bm S_\mathcal{G}$ & $\triangleq$ & $N \times N$ adjacency matrix\\	
		$\mathcal{\bm A}$ & $\triangleq$ & $N \times d$ matrix of node attributes\\
		$\mathbf{E}$ & $\triangleq$ & $N\times F$ matrix of embeddings\\
		$\mathbf{e}_i$ & $\triangleq$ & $F \times 1$ mbedding vector of node $v_i$ \\
		$a$ & $\triangleq$ & scalar \\
		$ \bm{a}$	& $\triangleq$ & vector \\		
		$\bm{A}$ & $\triangleq$ & matrix\\
		$\underline{\bm A}$ & $\triangleq$ & tensor \\
		$\bm{A}_k$ & $\triangleq$ & $k$-th frontal slab of tensor $\underline{\bm{A}}$\\		
		$\bm{A}^T$ & $\triangleq$ & transpose of matrix $\bm{A}$\\	
		$\lVert\bm{A}\rVert_F$ & $\triangleq$ & Frobenius norm of matrix $\bm{A}$ \\
		$\otimes$& $\triangleq$ & Kronecker product  of two matrices \\	
		$\odot$& $\triangleq$ & Khatri-Rao (columnwise Kronecker) product  \\
		$\lfloor x \rfloor$ & $\triangleq$ &  largest integer that is less than or equal to $x$\\
		$\bm I$& $\triangleq$ &  Identity matrix\\
		$\bm 1$& $\triangleq$ &  vector of ones\\
		\bottomrule
	\end{tabular}}
	\label{tab:TableOfNotationForMyResearch}
\end{table}
\vspace{-0.4cm}
\section{Preliminaries}\label{prelim}
Before moving into the core of the paper, we briefly discuss some tensor algebra preliminaries to facilitate the exposition. The reader is referred to \cite{sidiropoulos2017tensor,kolda2009tensor} for further background on tensors.

A third-order tensor $\underline{\bm X}\in\mathbb{R}^{I\times J\times K}$ is a three-way with elements $\underline{\bm X}(i,j,k)$. It comprises three modes -- columns $\underline{\bm X}(i,:,k)$ ($:$ stands for $\left\{1,\cdots,\text{end}\right\}$, where $\text{end}=J$ here), rows $\underline{\bm X}(:,j,k)$, and fibers $\underline{\bm X}(i,j,:)$; and three types of slabs -- horizontal $\underline{\bm X}(i,:,:)$, vertical $\underline{\bm X}(:,j,:)$ and frontal $\underline{\bm X}(:,:,k)$. A rank-one tensor $\underline{\bm Z}\in\mathbb{R}^{I\times J\times K}$ is the outer product of three vectors, $\bm a\in\mathbb{R}^{I},~\bm b \in\mathbb{R}^{J},~\bm c \in\mathbb{R}^{K}$, denoted as $\underline{\bm Z}=\bm a\circ \bm b\circ \bm c$, where $\circ$ is the outer product operator. 

Any third order tensor can be decomposed into a sum of three way outer products (rank one tensors), i.e. 
\vspace{-0.2cm}
\begin{equation}\label{PD2}
\underline{\bm X}(i,j,k)=\sum_{f=1}^F{\bm A}(i,f){\bm B}(j,f){\bm C}(k,f),
\end{equation}
where $\bm{A}=[\bm a_1,\dots,\bm a_F]\in\mathbb{R}^{I\times F},~\bm{B}=[\bm b_1,\dots,\bm b_F]\in\mathbb{R}^{J\times F},~\bm{C}=[\bm c_1,\dots,\bm c_F]\in\mathbb{R}^{K\times F}$ are the low-rank matrix factors. The minimum $F$ needed to synthesize $\underline{\bm X}$, is the \textit{rank} of tensor $\underline{\bm X}$ and the corresponding decomposition is known as the \textit{canonical polyadic decomposition} (CPD) of $\underline{\bm X}$ \cite{harshman1994parafac}, denoted as $\underline{\bm X}=\left\llbracket{\bm A},{\bm B},{\bm C}\right\rrbracket$.  
A striking property, that differentiates tensors from matrices, is that the CPD of a tensor is essentially unique under mild conditions, even if the rank is higher than the dimensions. A generic result on the uniqueness of the CPD follows.

\begin{Theorem}\label{thm:CPD_generic}
	\cite[p.~1019-1021]{chiantini2012generic} 
	Let $\underline{\bm X}=\left\llbracket{\bm A},{\bm B},{\bm C}\right\rrbracket$ with $\bm A : I\times F$, $\bm B : J\times F$, and $\bm C : K\times F$. Assume that ${\bm A}$, ${\bm B}$ and ${\bm C}$ are drawn from an absolutely continuous joint distribution with respect to the Lebesgue measure in $\mathbb{R}^{(I+J+K)F}$. Also assume $I\geq J\geq K$ without loss of generality. If $F \leq 2^{\lfloor\log_2 J\rfloor+\lfloor\log_2 K\rfloor-2}$, then the decomposition of $\underline{\bm X}$ in terms of $\bm A, \bm B$, and $\bm C$ is essentially unique, almost surely. 
\end{Theorem}

Here, essential uniqueness means that if $\tilde{\bm A},\tilde{\bm B},\tilde{\bm C}$ also satisfy $\underline{\bm X}=\llbracket \tilde{\bm A},\tilde{\bm B},\tilde{\bm C}\rrbracket$, then ${\bm A}=\tilde{\bm A}{\bm \Pi}{\bm \Lambda}_1$, ${\bm B}=\tilde{\bm B}{\bm \Pi}{\bm \Lambda}_2$, and ${\bm C}=\tilde{\bm C}{\bm \Pi}{\bm \Lambda}_3$, where ${\bm \Pi}$ is a permutation matrix and ${\bm \Lambda}_i$ is a full rank diagonal matrix such that ${\bm \Lambda}_1{\bm \Lambda}_2{\bm \Lambda}_3={\bm I}$.

A tensor can be represented in a matrix form using the \textit{matricization} operation. There are three common ways to matricize (or unfold) a third-order tensor, by stacking columns, rows, or fibers of the tensor to form a matrix. 
To be more precise let:
\begin{equation}
\bm{\underline{X}}(:,:,k)=\bm X_k\in\mathbb{R}^{I\times J},
\end{equation}
where $\bm X_k$ are the frontal slabs of tensor $\bm{\underline{X}}$ and in the context of this paper they model adjacency matrices, powers of the adjacency, node attributes or attribute similarity matrices.
Then the mode-$1$, mode-$2$ and mode-$3$ unfoldings of $\underline{\bm X}$ can be cast as:
\begin{equation}
\bm{{X}^{(1)}}=\begin{bmatrix}
\bm X_1, \bm X_2, \dots, \bm X_K
\end{bmatrix}^T=({\bm C}\odot{\bm B}){\bm A}^T\in\mathbb{R}^{JK\times I},
\end{equation}
\begin{equation}
\bm{{X}^{(2)}}=\begin{bmatrix}
\bm X_1^T, \bm X_2^T, \dots, \bm X_K^T
\end{bmatrix}^T=({\bm C}\odot{\bm A}){\bm B}^T\in\mathbb{R}^{IK\times J},
\end{equation}
\begin{equation}
\bm{{X}^{(3)}}=\begin{bmatrix}
\bm X_1(:,1), \bm X_2(:,1), \cdots, \bm X_K(:,1)\\
\vdots\\ \bm X_1(:,J), \bm X_2(:,J), \cdots, \bm X_K(:,J)
\end{bmatrix}=({\bm B}\odot{\bm A}){\bm C}^T\in\mathbb{R}^{IJ\times K}.
\end{equation}

\vspace{-0.2cm}

\section{Problem Statement}

We begin the discussion with the definition of node embedding. Let 
$\mathcal{G}:=\{ \mathcal{V},\mathcal{E}\}$ be a directed or undirected graph, with $\mathcal{V}$ being the set of $N=|\mathcal{V}|$ nodes, and $\mathcal{E} \subseteq \mathcal{V}\times \mathcal{V}$ being the set of edges. We are also given a set of attributes $\mathcal{\bm A}$ for each node. Node embedding aims to map each node to a vector in $F-$dimensional Euclidean space. Formally, the node embedding task seeks for a function $f(\cdot):\mathcal{G},\mathcal{A}\rightarrow\mathbb{R}^{N\times F}$, where $F\ll N$. The node embeddings can be represented by matrix $\bm E=\left[ \bm e_1, \bm e_2, \cdots, \bm e_N \right]^T$, where each row that contains the $F$-dimensional embedding of each node. 

\subsection{Related work}

Recent work \cite{al2018t} proposed building a tensor $\underline{\bm X}$ whose first frontal slab ${\bm X}_1$ is the network adjacency matrix, while its second frontal slab ${\bm X}_2$ is the an attribute adjacency, obtained by computing the set of $k$ nearest neighbors \cite{peterson2009k} of each node in attribute space. In other words, the attributes of a given node are viewed as a vector in Euclidean space, and the $k$ closest attribute vectors of other nodes in the network are used to define the neighbors of the given node.  The number of nearest neighbors is a parameter that needs to be tuned. A second adjacency matrix is produced this way, which however is not necessarily symmetric (even if the original network adjacency is). Joint analysis of these two adjacency matrices yields embeddings that reflect both pieces of information -- but are not geometry-preserving, because (approximately) reproducing these adjacency matrices has no geometric motivation / interpretation.

In this work we propose a principled formulation that directly aims to produce an embedding that can reproduce the distances between nodes in terms of their network adjacency and in terms of their attributes. We find common latent dimensions that explain both sets of distances. With proper weighting, the resulting embedding vectors reproduce the adjacency distances; with another weighting, they reproduce the attribute distances (and these weights are a by-product of our analysis). Either way, the latent dimensions are derived from (and reflect) both sets of distances. This is why we call the approach {\em geometry preserving}. Depending on the downstream task, different weighting schemes would be more appropriate. Our formulation draws from multi dimensional scaling (MDS), which is briefly reviewed next.

\subsection{Multi dimensional scaling}
MDS is a distance-preserving mapping, visualization, and embedding tool \cite{torgerson1952multidimensional,kruskal1978multidimensional,schiffman1981introduction,cox2008multidimensional}. Given an $N \times N$ matrix $\bm D$ of distances between $N$ entities, MDS seeks to find $N$ points in low-dimensional space (typically 2- or 3-dimensional, for visualization purposes) that approximately exhibit the given distances. Various distances (and pseudo-distances) can be used for MDS. The most popular is the Euclidean distance, leading to the classical MDS, but there exist non-metric versions of MDS which seek to preserve ordering as opposed to distances \cite{kruskal1978multidimensional}. We next briefly review classical MDS. Let $\bm D^{(2)}\in\mathbb{R}^{N\times N}$ be the matrix of squared distances between $N$ entities, with $\bm D^{(2)}(i,j)$ being the squared distance between entity $i$ and entity $j$. Now let $\bm e_i$ be the vector representation of entity $i$ in a low $F$-dimensional Euclidean space. Then it holds that:
\begin{equation}
    \bm D^{(2)}(i,j) = \lVert \bm e_i -\bm e_j\rVert^2 = \lVert \bm e_i \rVert^2+\lVert \bm e_j\rVert^2 -2 \bm e_i^T \bm e_j
\end{equation}
Since the objective is to learn the $\{\bm e_i\}_{i=1}^N$ we would like to end up with an expression that ignores the squared norms $\lVert \bm e_i \rVert^2,\lVert \bm e_j\rVert^2$ and will be easy to factor. In this direction we observe that:
\begin{equation}\label{eq:mds2}
    \bm D^{(2)} = \bm g \bm 1^T + \bm 1 \bm g^T -2\bm E\bm E^T,
\end{equation}
where $\bm g = \left[ \bm e_1^T\bm e_1, \dots, \bm e_N^T\bm e_N \right]^T$. Double centering both sides yields:
\begin{align}\label{eq:mds3}
   &\left(\bm I-\frac{1}{N}\bm 1 \bm 1^T\right) \bm D^{(2)}\left(\bm I-\frac{1}{N}\bm 1 \bm 1^T\right) =\left(\bm I-\frac{1}{N}\bm 1 \bm 1^T\right)\bm g \bm 1^T\left(\bm I-\frac{1}{N}\bm 1 \bm 1^T\right)+\nonumber\\& \left(\bm I-\frac{1}{N}\bm 1 \bm 1^T\right)\bm 1 \bm g^T\left(\bm I-\frac{1}{N}\bm 1 \bm 1^T\right) -\bm \left(\bm I-\frac{1}{N}\bm 1 \bm 1^T\right)2\bm E \bm E^T\left(\bm I-\frac{1}{N}\bm 1 \bm 1^T\right),
\end{align}
which is equivalent to
\begin{equation}\label{mdsNikos}
-\frac{1}{2}\left(\bm I-\frac{1}{N}\bm 1 \bm 1^T\right) \bm D^{(2)}\left(\bm I-\frac{1}{N}\bm 1 \bm 1^T\right)= \bm E\bm E^T,
\end{equation}
since $\left(\bm I-\frac{1}{N}\bm 1 \bm 1^T\right)\bm 1= 0$ and we can assume without loss of generality that matrix $\bm E$ is already centered.
The solution for $\bm E$ is given by
\begin{equation}\label{eq:MDS_solution}
\bm E = \bm U \sqrt{\bm\Lambda_F},
\end{equation}
where $\bm U\in\mathbb{R}^{N\times F}$ is the matrix of $F$ principal eigenvectors and $\bm \Lambda_F \in\mathbb{R}^{F\times F}$ a diagonal matrix with the $F$ principal eigenvalues of $-\frac{1}{2}\left(\bm I-\frac{1}{N}\bm 1 \bm 1^T\right) \bm D^{(2)}\left(\bm I-\frac{1}{N}\bm 1 \bm 1^T\right)$. In the non-ideal case where $\bm D^{(2)}$ is inexact, assuming that the left hand side of (\ref{mdsNikos}) remains (or is projected to be) positive semidefinite, \eqref{eq:MDS_solution} gives the best vector representation of the entities in an $F$-dimensional space {\em after double-centering}, albeit that is not optimal from the viewpoint of preserving the original distances. For the latter, we need to resort to iterative algorithms that minimize a suitable cost (or {\em stress}) function, but that is often not necessary in practice.  

MDS has also been generalized to the case where more than one distance matrices are available for a set of entities \cite{carroll1970analysis}. For example, the entities could be a set of $N$ different products and $K$ individuals are asked to rate their similarity or dissimilarity. This results in $K$ different $N \times N$ distance matrices for the $N$ products. To be more precise let $\bm D^{(2)}_k\in\mathbb{R}^{N\times N}$ be the $k$-th given distance matrix. Three-way MDS forms a third-order tensor $\underline{\bm X}\in\mathbb{R}^{N\times N\times K}$ as:
\begin{equation}
    \underline{\bm X}(:,:,k) = \left(\bm I-\frac{1}{N}\bm 1 \bm 1^T\right) \bm D^{(2)}_k\left(\bm I-\frac{1}{N}\bm 1 \bm 1^T\right)
\end{equation}
and performs CPD of $\underline{\bm X}$ to find a joint $F$-dimensional representation of the entities.

\subsection{GAGE: Geometry preserving Attributed Graph Embeddings}
In the previous section we introduced the task of unsupervised node embedding. The objective of this task is to map each node of the network to a low dimensional vector representation in the Euclidean space. It is desirable that the low dimensional embedding contains as much connectivity and attribute information as possible and progress in this direction is the key to successful embeddings. 

In this section, motivated by the benefits of MDS, we propose a novel unsupervised node embedding scheme that works with attributed networks. The proposed node embedding scheme attempts to preserve the network geometry inferred both from connectivity and attribute information. Furthermore, the node embeddings are unique. Note that uniqueness is a fundamental property that each embedding should enjoy. It offers a unique representation of each node, which is necessary for any form of interpretability and also guarantees that the embedding is permutation invariant. In other words any permuted version of the adjacency yields the same embedding for each node. Finally the proposed representation model is flexible in the sense that it can handle both directed and undirected graphs and does not require connectivity and attributed information for every node. In other words embeddings can be produced for nodes with either missing connectivity information or missing attributes.

Traditional MDS starts from a distance matrix and looks for vector representations of the nodes. In our setting, we are given the adjacency representation of each node along with a vector of attributes. The obvious approach would be to try and learn low-dimensional node embeddings directly from the high-dimensional graph and attribute representation of each node. However, since our objective is the produced embeddings to preserve the network geometry in terms of Euclidean distances, we propose to follow a different route. In particular, given the adjacency and the attributes of the network we compute distance matrices, one for the connectivity information and another for the attribute information. This transformation from adjacency and attributes to connectivity and attribute distances is the key to our proposed geometry preserving embeddings.
Then we decompose the tensor of distances, using the CPD model, and produce the low-dimensional embeddings. As we see later in the section, the produced embeddings, which are formed from the CPD factors, can reproduce both the connectivity and attribute distances. Note that, from a computational viewpoint, instantiating the Euclidean distance matrices of connectivity and attributes might be prohibitive, since it destroys the sparsity structure. Interestingly, there is a elegant way to overcome it.

In order to facilitate the analysis let ${\bm S}_{\mathcal{G}}\in\{0,1\}^{N\times N}$ denote the adjacency matrix of graph $\mathcal{G}$ and $\mathcal{\bm A}\in\mathbb{R}^{N\times d}$ be the matrix the contains in row $i$ $d$ attributes or features of vertex $i$. Also let $\bm Y_1 = \bm S_{\mathcal{G}}$ and $\bm Y_2 =\mathcal{\bm A}$. Taking a closer look at equation \eqref{eq:mds3} we observe that double centering the matirx of Euclidean distances between the rows of $\bm Y_1$ or $\bm Y_2$ is equal to double centering $\bm Y_1\bm Y_1^T$ or $\bm Y_2\bm Y_2^T$. This is due to the fact that $\bm Y_1$ or $\bm Y_2$ contain the generating vectors of the distance matrices and equation \eqref{eq:mds2} always holds. 
We now transform the adjacency and attribute to distance matrices: 

\vspace{-0.2cm}
\begin{align}\label{eq:MDS}
\bm X_1 = \left(\bm I-\frac{1}{N}\bm 1 \bm 1^T\right) \bm Y_1\bm Y_1^T\left(\bm I-\frac{1}{N}\bm 1 \bm 1^T\right),\\    
\bm X_2 = \left(\bm I-\frac{1}{N}\bm 1 \bm 1^T\right) \bm Y_2\bm Y_2^T\left(\bm I-\frac{1}{N}\bm 1 \bm 1^T\right).
\end{align}
 Note that $\bm X_1(i,j)$ denotes the squared Euclidean distance (after double centering) between $\bm S_G(i,:)$ and $\bm S_G(j,:)$, i.e., two rows of the adjacency matrix. Also $\bm X_2(i,j)$ denotes the squared Euclidean distance (after double centering) between $\mathcal{\bm A}(i,:)$ and $\mathcal{\bm A}(j,:)$, i.e., two attributed information of different nodes. It is important to mention that in most applications $\bm S_{\mathcal{G}}$ and $\mathcal{\bm A}$ are sparse matrices which facilitates storage and computation requirements. Double centering these matrices automatically yields dense matrices. However, as we will see next our approach doesn't instantiate the dense $\bm X_1$ and $\bm X_2$ but works with sparse $\bm Y_1$ and $\bm Y_2$, which is crucial to keep the computational and memory complexity of the algorithm low.

To compute the node embeddings of the attributed network, we propose to employ the following optimization scheme:
\begin{align}\label{eq:emb2}
\min_{\bm U,\bm\Lambda_1,\bm\Lambda_2} \lVert {\bm X}_1-\bm U\bm\Lambda_1\bm U^T\rVert_F^2+\lVert {\bm X}_2-\bm U\bm\Lambda_2\bm U^T\rVert_F^2,
\end{align}
where $\bm U\in\mathbb{R}^{N\times F}$ and $\bm\Lambda_1,\bm\Lambda_2$ are real and positive valued $F\times F$ diagonal matrices. Problem \eqref{eq:emb2} is the rank $F$ CPD of tensor $\underline{\bm X}\in\mathbb{R}^{N\times N\times 2}$, with frontal slabs $\underline{\bm X}(:,:,1)=\bm X_1$ and $\underline{\bm X}(:,:,2)={\bm X_2}$. The CPD model for $\underline{\bm X}$ takes the form:
\begin{equation}
\bm{\underline{X}}=\llbracket {\bm U},{\bm U},{\bm C}\rrbracket, ~~~~\bm{C}(i,:)^T=\text{diag}\left(\bm\Lambda_i\right),~i=1,2,
\end{equation}
where $\text{diag}\left(\bm\Lambda_i\right)$ is the diagonal vector of $\bm\Lambda_i$. The proposed embedding for vertex $i$ is :
\begin{equation}
   \bm e_i= \bm E(i,:)^T=\text{diag}\left(\sqrt{\lambda\bm C(:,1)^T+(1-\lambda)\bm C(:,2)^T}\right)~\bm U(i,:)^T,
\end{equation}
where $\text{diag}\left(\sqrt{\lambda\bm C(:,1)^T+(1-\lambda)\bm C(:,2)^T}\right)$ gives the diagonal matrix of vector $\sqrt{\lambda\bm C(:,1)^T+(1-\lambda)\bm C(:,2)^T}$. Note that the $0\leq\lambda\leq 1$ parameter balances the contribution of each distance measure (connectivity or attribute) in the final embedding. For $\lambda =1$ the focus is completely on the connectivity distances, whereas for $\lambda =0$ the emphasis is on the attribute distances.

Invoking the uniqueness properties o the CPD (see Theorem 1 for details) we have shown the following result:
\begin{Result}
If tensor $\underline{\bm X}$ has indeed low-rank, $F$, there exist vectors in $F$ dimensional space that generate the given sets of distances (with appropriate weights). Then the $\texttt{GAGE}$ embeddings for the correct $F$ are unique, permutation invariant and will exactly reproduce both sets of distances for $\lambda =0$ and $\lambda =1$.
\end{Result}
The above result also implies that embeddings of dimension less than $F$ cannot reconstruct the set of distances and embeddings of dimension larger than $F$ are not unique. 

\section{Algorithmic framework}
\label{sec:Algorithms}
In this section we discuss the algorithmic aspects of our approach.
\subsection{The \texttt{GAGE} algorithm}
The computation of the proposed node embeddings boils down to solving the problem in \eqref{eq:emb2}. This is a CPD problem of an $N\times N\times 2$ tensor with a special sparsity structure on the frontal slabs. CPD computation is a non-convex optimization problem and in general NP-hard. However, exact CPD can be reduced to eigenvalue decomposition (EVD) in certain cases -- notably when tensor rank is low enough \cite{sanchez1990tensorial,domanov2014canonical}. Such an approach is not guaranteed to produce the optimal solution, but it often works well in practice, and it also serves as good initialization for more sophisticated optimization approaches. Developing a computationally efficient algebraic initialization approach to tackle the problem in \eqref{eq:emb2} is therefore an important pivot for the proposed algorithm. This is \texttt{GAGE-EVD}, which is summarized in Algorithm \ref{algo:gage_evd}. The first step of the approach is to form the doubly centered frontal slabs. Note that instantiating $\bm X_1,~\bm X_2$ is not required and we can directly work with $\bm Y_1,~\bm Y_2$ as shown in Appendix \ref{app:GAGE-evd}; \texttt{GAGE-EVD} exploits sparsity and the special problem structure to mitigate memory and complexity requirements. The next step is to compute the $F$ principal eigenvectors $\bm V$ of $\bm X_1^T\bm X_1+\bm X_2^T\bm X_2$. Towards this, end we employ the orthogonal iterations method \cite{golub2013matrix} which also exploits the special sparsity structure to enable lightweight computations. Finally, we form $\bm S_1 = \bm V^T\bm X_1 \bm V,~\bm S_2 = \bm V^T\bm X_2 \bm V$ which are dense but small ($F\times F$) matrices and compute the eigenvalue decomposition of $\bm S_2\bm S_1^{-1}$. Then $\bm U$ is computed as $\bm U^T = \tilde{\bm U}^{-1}\bm S_1$. In terms of computational complexity, the main bottleneck of \texttt{GAGE-EVD} is computing the EVD of $\bm X_1^T\bm X_1+\bm X_2^T\bm X_2$. Using the orthogonal iterations method, this EVD can be computed efficiently in $\mathcal{O}({NF^2})$ flops. The remaining operations involve $F\times F$ matrices and are computationally light. Detailed description of the algorithmic updates along with computational complexity and memory requirements is given in Appendix \ref{app:GAGE-evd}.

After computing an initial estimate of matrix $\bm U$, we feed it to the main \texttt{GAGE} algorithm, which is summarized in Algorithm \ref{algo:gage}. To tackle the problem in \eqref{eq:emb2} \texttt{GAGE} follows an alternating least squares approach, with the first two factors $\bm U, \bm U^{'}$ not constrained to be equal (see Algorithm \ref{algo:gage}). In each update, we fix two factors and solve for the remaining one. We repeat this procedure in an alternating fashion. The update for each step is a linear system of equations and can be solved efficiently without instantiating the dense tensor $\underline{\bm X}$ or any of the Khatri-Rao products, i.e., $(\bm C\odot\bm U^{'}),~(\bm C\odot\bm U^{'}),~(\bm U^{'}\odot\bm U)$. The details are presented in Appendix \ref{app:GAGE}. Note that due to the algebraic initialization, the \texttt{GAGE} algorithm converges in only a few steps (usually fewer than 10) in our experiments.
\setlength{\textfloatsep}{0.1cm}
\begin{algorithm}[t]
	\caption{\texttt{GAGE-EVD}}
	\label{algo:gage_evd}
	\begin{algorithmic}
	\small{
		\STATE {\bfseries Input:} $\bm Y_1=\bm{S}_{\mathcal{G}},\bm Y_2 =\mathcal{\bm A},~F $.
		\STATE {\bfseries Output:} $\bm{U}$.
		\STATE $\bm X_1 = \left(\bm I-\frac{1}{N}\bm 1 \bm 1^T\right) \bm Y_1\bm Y_1^T\left(\bm I-\frac{1}{N}\bm 1 \bm 1^T\right);$
		\STATE $\bm X_2 = \left(\bm I-\frac{1}{N}\bm 1 \bm 1^T\right) \bm Y_2\bm Y_2^T\left(\bm I-\frac{1}{N}\bm 1 \bm 1^T\right);$\\
        \STATE $\bm V\bm\Sigma\bm V^T \leftarrow$ EVD$\left(\bm X_1^T\bm X_1+\bm X_2^T\bm X_2,F\right);$
		\STATE $\bm S_1 = \bm V^T\bm X_1 \bm V,~\bm S_2 = \bm V^T\bm X_2 \bm V;$
		\STATE $\tilde{\bm U} \leftarrow$ EVD $\left(\bm S_2\bm S_1^{-1}\right);$
		\STATE $\bm U^T = \tilde{\bm U}^{-1}\bm V^T;$}
	\end{algorithmic}
\end{algorithm}
\setlength{\floatsep}{0.1cm}
\begin{algorithm}[t]
	\caption{\texttt{GAGE}}
	\label{algo:gage}
	\begin{algorithmic}
	\small
		\STATE {\bfseries Input:} $\bm Y_1=\bm{S}_{\mathcal{G}},\bm Y_2 =\mathcal{\bm A},~\bm{U}$.
		\STATE {\bfseries Output:} $\bm{E}$.
		\STATE $\bm X_1 = \left(\bm I-\frac{1}{N}\bm 1 \bm 1^T\right) \bm Y_1\bm Y_1^T\left(\bm I-\frac{1}{N}\bm 1 \bm 1^T\right);$
		\STATE $\bm X_2 = \left(\bm I-\frac{1}{N}\bm 1 \bm 1^T\right) \bm Y_2\bm Y_2^T\left(\bm I-\frac{1}{N}\bm 1 \bm 1^T\right);$\\
		\STATE $\underline{\bm{X}}(:,:,1) = \bm X_1,~\underline{\bm{X}}(:,:,2) = \bm X_2;$
		\STATE$\bm C \leftarrow$ solve $\bm{{X}^{(3)}}=(\bm U\odot\bm U)\bm C^T;$
		\STATE$\bm U^{'}=\bm U$;
		\REPEAT
		\STATE $\bm U \leftarrow$ solve $\bm{{X}^{(1)}}=(\bm C\odot\bm U^{'})\bm U^T;$
		\STATE $\bm U^{'} \leftarrow$ solve $\bm{X}^{(2)}=(\bm C\odot\bm U)\bm U^{'^T};$
		\STATE $\bm C \leftarrow$ solve $\bm{X}^{(3)}=(\bm U^{'}\odot\bm U)\bm C^T;$
		\UNTIL convergence
		\STATE $\bm E = \bm U~\text{diag}\left(\sqrt{\lambda\bm C(:,1)^T+(1-\lambda)\bm C(:,2)^T}\right);$
	\end{algorithmic}
\end{algorithm}
\setlength{\floatsep}{0.1cm}
\vspace{-0.2cm}
\section {Experiments}\label{sec:num}
In this section we demonstrate the performance of the proposed algorithmic framework and showcase its effectiveness in experiments with real attributed network data. All algorithms were implemented in Matlab or Python, and executed on a Linux server comprising 8 cores at 3.6GHz with 32GB RAM.
\vspace{-0.3cm}
\subsection{Data}
We used the following real-world networks (see also Table 2). 

\begin{itemize}


\item \textbf{\texttt{BlogCatalog}}. A social network of bloggers in BlogCatalog platform. Each blogger uses several keywords to describe their blogs. These keywords have been used as attributes for the node-bloggers. There are 6 different classes of bloggers according to the content of their blogs. The attributes dimension represents the dictionary and each node is encoded with a sparse bag-of-words representation. 

\item \textbf{\texttt{WebKB}} \cite{getoor2005link}. A network of webpages from computer science departments categorized into 5 topics: faculty, student, project, course, other. The attributes dimension is a dictionary of words that appear in the webpages.
	

\item \textbf{\texttt{Wikipedia}} \cite{yang2015network}. A network of documents and their Wikipedia links. The documents are grouped into 19 classes and the attribute information corresponds to sparse TFIDF features.
			
\end{itemize}
\begin{table*}
  \caption{Datasets}
  \label{tab:freq}
  \scalebox{0.9}{\begin{tabular}{ccccccc}
    \toprule
    Dataset&\# Vertices& \# Edges& Attribute dimension & \# Classes & Network Type & Feature Type\\
    \midrule
    \texttt{Wikipedia} & 2,405& 23,192 & 4,973 & 19 & Language & Text associated info\\
    \texttt{WebKB} & 877& 2,776 & 1,703 & 5 & Citation & Unique words\\
    \texttt{BlogCatalog} & 5,196& 686,972 & 8,189 & 6 & Social & Keywords\\
  \bottomrule
\end{tabular}}
\end{table*}
\vspace{-0.3cm}
\subsection{Baselines}

\begin{itemize}

\item \textbf{Deepwalk} \cite{perozzi2014deepwalk}. \texttt{Deepwalk} generates  truncated random walks from each node, to learn low dimensional representations of nodes using a SkiGram model. We set the number of walks per node $\gamma= 80$, walk length $t= 40$ and window size $w= 10$ as suggested in \cite{perozzi2014deepwalk}. This method does not use the attributes, and it is not expected to work as well as the other methods that do. We include it, since it remains a strong contender, and as a means to gauge the improvement afforded by having access to the node attributes.    
	
\item \textbf{T-Pine} \cite{al2018t}. 
A tensor factorization based approach. The first frontal slab is the adjacency of the graph and the second frontal slab is the a k nearest neighbor matrix computed using the distances between the node attributes. The k nearest neighbor parameter is set to $k=8$ for \texttt{Wikipedia}, $k=40$ for \texttt{WebKB} as suggested in \cite{al2018t} and $k=50$ for \texttt{BlogCatalog}. 
\item \textbf{Graph-AE} \cite{kipf2016variational}. A graph convolutional network (GCN) generalization for unsupervised node embedding. Graph-AE uses a (GCN) encoder and a simple inner product decoder.

\item \textbf{Graph-VAE} \cite{kipf2016variational}. A variational auto encoder (VAE) alternative to \texttt{Graph-AE}. Both \texttt{Graph-AE} and \texttt{Graph-VAE} are trained using $200$ epochs and $0.01$ learning rate. The dimension of the hidden layer is twice the number of the embedding dimension. We use $5\%$ of the data for validation.

\item \textbf{TADW} \cite{yang2015network}. Text associated Deepwalk (TADW) employs a matrix factorization framework to learn network representations using the adjacency matrix as well as textual information features.

\item \textbf{DGI} \cite{velickovic2019deep}. Deep Graph Infomax (DGI) uses a graph convolutional neural network architecture to learn node embeddings for attributed networks in an unsupervised manner. We train for maximum $1000$ epochs using the code provided by the authors and set the `patience' parameter equal to $20$ and learning rate equal to $0.001$, as suggested in \cite{velickovic2019deep}.
\item \textbf{AGE} \cite{cui2020adaptive}. Adaptive Graph Encoder for Attributed Graph Embedding (AGE) uses a Laplacian smoothing filter along with an adaptive encoder to perform attributed node embedding. We use $400$ epochs and learning rate equal to $10^{-3}$ for training, as suggested in the author's code.
\item \textbf{DANE} \cite{gao2018deep}. Deep Attributed Network Embedding (DANE) adopts a 2-branch encoder-decoder architecture to learn attributed node embeddings. The first branch is associated with the connectivity information of the network, whereas the second one utilizes the attribute information. We use $500$ epochs to train the autoencoder with learning rate and dropout probability equal to $10^{-5}$ and $0.2$ respectively.
\end{itemize}
For all baselines we use the publicily available code provided by the authors.
\vspace{-0.2cm}
\subsection{Node classification}
We first test the performance of the proposed \texttt{GAGE} along with the baselines in a node classification task. The procedure is divided in two steps. In the first step the algorithms learn the node embeddings in a unsupervised manner, i.e., without using label information. In the second step the labels along with the learned embeddings are split into training and testing sets. Then the training data are fed to a one-versus-all logistic regression classifier with $l_2$ norm regularization. We test 3 different training-testing splits, i.e., 0.9-0.1, 0.5-0.5, and 0.1-0.9 and run 10 shuffles for each split. To assess the performance of the competing algorithms we measure the average micro and macro F1 score for 2 different embedding dimensions. For the \texttt{GAGE} embeddings we set $\lambda = 0.8$. The results for the three different datasets are presented in Tables \ref{tab:Results wiki}, \ref{tab:Results WebKB}, \ref{tab:Results BlogCatalog}.

\begin{table*}
	\caption{Average score and standard deviation over 10 shuffles for \texttt{Wikipedia}}
	\label{tab:Results wiki}
    \scalebox{0.8}{
	\begin{tabular}{ccccccccc}
		\toprule
		Algorithm &  dimension &  micro (0.9) & macro (0.9)& micro (0.5) & macro (0.5)& micro (0.1) & macro (0.1)\\
		\midrule
		\multirow{2}{*}{ \texttt{GAGE}} 
		&{{64}}   & $ \mathbf{0.7402 \pm 0.0308} $ & $ \mathbf{0.5331 \pm 0.0239} $ & $ \mathbf{0.7303 \pm 0.0125} $ & $ \mathbf{0.5262 \pm 0.0217} $ & $ \mathbf{0.6309 \pm 0.0217} $ & $ \mathbf{0.423 \pm 0.0246} $\\
		&{{128}}   & $ 0.7656 \pm 0.0255 $ & $ \mathbf{0.5924 \pm 0.0337} $ & $ 0.736 \pm 0.0104 $ & $ \mathbf{0.5802 \pm 0.0198} $& $ \mathbf{0.649 \pm 0.0179} $ & $ \mathbf{0.4728 \pm 0.0261} $\\
		\hline
		\multirow{2}{*}{ \texttt{T-PINE}} 
		
		&{{64}}   & $ 0.6788 \pm 0.0312 $ & $ 0.4039 \pm 0.0158 $ & $ 0.6619 \pm 0.0072 $ & $ 0.3949 \pm 0.0047 $&$ 0.5912 \pm 0.0139 $ & $ 0.3535 \pm 0.0042 $ \\
		&{{128}} & $ \mathbf{0.766 \pm 0.0234} $ & $ 0.523 \pm 0.0183 $ & $ \mathbf{0.745 \pm 0.009} $ & $ 0.5069 \pm 0.0121 $ & $ 0.6364 \pm 0.0081 $ & $ 0.4205 \pm 0.0144 $ \\
		\hline
		\multirow{2}{*}{\texttt{Deepwalk}} 
		&{{64}}   & $ 0.6177 \pm 0.0309 $ & $ 0.3632 \pm 0.0213 $& $ 0.6136 \pm 0.0038 $ & $ 0.3736 \pm 0.0119 $& $ 0.5773 \pm 0.0084 $ & $ 0.3415 \pm 0.0126 $\\
		&{{128}}   & $ 0.6236 \pm 0.0333 $ & $ 0.362 \pm 0.0175 $& $ 0.614 \pm 0.006 $ & $ 0.3731 \pm 0.0111 $& $ 0.5811 \pm 0.0095 $ & $ 0.3444 \pm 0.0126 $\\
		\hline
		\multirow{2}{*}{\texttt{Graph-AE}} 
		&{{64}}  & $ 0.6759 \pm 0.0314 $ & $ 0.4512 \pm 0.0335 $& $ 0.6481 \pm 0.0117 $ & $ 0.4254 \pm 0.0193 $& $ 0.5669 \pm 0.0075 $ & $ 0.3452 \pm 0.0121 $\\
		&{{128}}   & $ 0.6747 \pm 0.0372 $ & $ 0.4327 \pm 0.0346 $& $ 0.6584 \pm 0.0082 $ & $ 0.4287 \pm 0.0203 $ & $ 0.5773 \pm 0.01 $ & $ 0.3536 \pm 0.0115 $\\
		\hline
		\multirow{2}{*}{\texttt{Graph-VAE}} 
		&{{64}}   & $ 0.6283 \pm 0.03 $ & $ 0.4108 \pm 0.0297 $& $ 0.6069 \pm 0.009 $ & $ 0.3804 \pm 0.0137 $& $ 0.5592 \pm 0.0112 $ & $ 0.3323 \pm 0.0124 $\\
		&{{128}}   & $ 0.67 \pm 0.0394 $ & $ 0.436 \pm 0.028 $& $ 0.6404 \pm 0.0119 $ & $ 0.4098 \pm 0.0195 $& $ 0.5742 \pm 0.0107 $ & $ 0.3431 \pm 0.0109 $\\
		\hline
		\multirow{2}{*}{\texttt{TADW}} 
		&{{64}}   & $0.7008 \pm0.0258$& $0.4541\pm0.0244$& $0.6990\pm0.0142$& $0.4781\pm0.0156$&$0.6160\pm0.0121$&$0.3996 \pm0.0126$\\
		&{{128}}   & $0.7510 \pm 0.0345$& $0.5378\pm0.0373$& $0.7168\pm0.0135$& $0.5170\pm0.0224$& $0.6309\pm0.0159$& $0.4221\pm 0.0218$\\
		\hline
		\multirow{2}{*}{\texttt{DGI}} 
		&{{64}}   &$ 0.5523 \pm 0.0258 $ & $ 0.2806 \pm 0.0095 $ &$ 0.4927 \pm 0.0182 $ & $ 0.2271 \pm 0.0124 $ &$ 0.3157 \pm 0.0323 $ & $ 0.0741 \pm 0.0139 $\\
		&{{128}}   &$ 0.5299 \pm 0.0254 $ & $ 0.252 \pm 0.0139 $ &$ 0.4563 \pm 0.0171 $ & $ 0.1825 \pm 0.0119 $ & $ 0.2924 \pm 0.0458 $ & $ 0.066 \pm 0.0148 $\\
		\hline
		\multirow{2}{*}{\texttt{AGE}} 
		&{{64}}   & $ 0.6996 \pm 0.0228 $ & $ 0.4398 \pm 0.009 $& $ 0.6826 \pm 0.0138 $ & $ 0.4261 \pm 0.0124 $& $ 0.6038 \pm 0.0217 $ & $ 0.3448 \pm 0.0258 $\\
		&{{128}}   & $ 0.7058 \pm 0.0323 $ & $ 0.452 \pm 0.0182 $& $ 0.6909 \pm 0.0094 $ & $ 0.437 \pm 0.0084 $& $ 0.5833 \pm 0.0217 $ & $ 0.3105 \pm 0.0254 $\\
		\hline
        \multirow{2}{*}{\texttt{DANE}} 
		&{{64}}   & $ 0.5029 \pm 0.0383 $ & $ 0.2575 \pm 0.0314 $& $ 0.4259 \pm 0.0243 $ & $ 0.1926 \pm 0.0134 $& $ 0.2409 \pm 0.025 $ & $ 0.0575 \pm 0.0124 $\\
		&{{128}}   & $ 0.6734 \pm 0.0339 $ & $ 0.4141 \pm 0.022 $& $ 0.6565 \pm 0.0112 $ & $ 0.3976 \pm 0.0093 $& $ 0.5321 \pm 0.0208 $ & $ 0.2706 \pm 0.0219 $ \\
		\hline
		\bottomrule
	\end{tabular}}
\end{table*}
\begin{table*}
	\caption{Average score over 10 shuffles for \texttt{WebKB}}
	\label{tab:Results WebKB}
	\scalebox{0.8}{
	\begin{tabular}{ccccccccc}
		\toprule
		Algorithm & dimension  & micro (0.9) & macro (0.9)& micro (0.5) & macro (0.5)& micro (0.1) & macro (0.1)\\
		\midrule
	\multirow{2}{*}{ \texttt{GAGE}} 

		&{{64}}   & $ \mathbf{0.8852 \pm 0.0375} $ & $ \mathbf{0.7588 \pm 0.0506} $& $ \mathbf{0.8547 \pm 0.0221} $ & $ \mathbf{0.7005 \pm 0.0228} $& $ \mathbf{0.7722 \pm 0.0233} $ & $ \mathbf{0.5701 \pm 0.0352} $\\
		&{{128}}   & $ \mathbf{0.8864 \pm 0.037} $ & $ \mathbf{0.7618 \pm 0.0645} $& $ \mathbf{0.8601 \pm 0.0148} $ & $ \mathbf{0.7024 \pm 0.0264} $& $ \mathbf{0.7566 \pm 0.0256} $ & $ \mathbf{0.5419 \pm 0.0372} $\\
		\hline
		\multirow{2}{*}{ \texttt{T-PINE}} 
		&{{64}}   & $0.8148\pm0.0318$ & $0.6504\pm0.0633$& $0.8016\pm0.0192$&$0.6361\pm0.0224$ &$0.7033\pm0.018$&$0.5204\pm0.0185$\\
		&{{128}}   & $0.7989\pm0.0297$ & $0.6394\pm0.0632$& $0.7743\pm0.0144$& $0.6141\pm0.0241$& $0.681\pm0.0201$& $0.4822\pm0.0166$\\
		\hline
		\multirow{2}{*}{\texttt{Deepwalk}} 
		&{{64}}   & $ 0.5081 \pm 0.0543 $ & $ 0.2627 \pm 0.0284 $& $ 0.4786 \pm 0.0206 $ & $ 0.2448 \pm 0.0217 $& $ 0.4367 \pm 0.0152 $ & $ 0.2228 \pm 0.0175 $\\
		&{{128}}  & $ 0.4977 \pm 0.049 $ & $ 0.2914 \pm 0.0437 $& $ 0.4674 \pm 0.0207 $ & $ 0.2487 \pm 0.0242 $& $ 0.4447 \pm 0.015 $ & $ 0.2249 \pm 0.0177 $\\
		\hline
	\multirow{2}{*}{\texttt{Graph-AE}} 
		&{{64}}   & $ 0.4591 \pm 0.0306 $ & $ 0.1261 \pm 0.0061 $& $ 0.4722 \pm 0.0144 $ & $ 0.1294 \pm 0.0029 $ &$ 0.4767 \pm 0.0079 $ & $ 0.1373 \pm 0.0119 $\\
		&{{128}}  & $ 0.4591 \pm 0.0306 $ & $ 0.1257 \pm 0.0058 $& $ 0.4715 \pm 0.0146 $ & $ 0.1281 \pm 0.0027 $ &$ 0.4732 \pm 0.0056 $ & $ 0.1285 \pm 0.001 $\\
		\hline
		\multirow{2}{*}{\texttt{Graph-VAE}} 
		&{{64}}  & $ 0.5261 \pm 0.0322 $ & $ 0.2435 \pm 0.0275 $& $ 0.5276 \pm 0.0135 $ & $ 0.2502 \pm 0.0123 $ &$ 0.4985 \pm 0.0165 $ & $ 0.2483 \pm 0.0267 $\\
		&{{128}} & $ 0.542 \pm 0.0446 $ & $ 0.2489 \pm 0.0206 $& $ 0.5376 \pm 0.0207 $ & $ 0.2521 \pm 0.012 $ &$ 0.5009 \pm 0.0185 $ & $ 0.2434 \pm 0.0258 $\\
		\hline
		\multirow{2}{*}{\texttt{TADW}} 
		&{{64}}   & $0.6931\pm 0.0344$& $0.5368\pm0.0562$& $0.6646\pm0.0226$& $0.4887\pm0.0355$& $0.5988\pm0.0190$& $0.3889\pm 0.0250$\\
		&{{128}}   & $0.7511\pm 0.0404$ & $0.6176\pm 0.0849$& $0.7200\pm 0.0252$& $0.5539\pm 0.0305$& $0.6287\pm 0.0213$& $0.4197\pm 0.0306$\\
		\hline
		\multirow{2}{*}{\texttt{DGI}} 
		&{{64}}   & $ 0.4705 \pm 0.0378 $ & $ 0.147 \pm 0.0199 $& $ 0.4797 \pm 0.0163 $ & $ 0.1444 \pm 0.0067 $&$ 0.4762 \pm 0.0068 $ & $ 0.1336 \pm 0.0036 $\\
		&{{128}}   & $ 0.4727 \pm 0.0374 $ & $ 0.1472 \pm 0.0205 $& $ 0.4772 \pm 0.0146 $ & $ 0.1378 \pm 0.0047 $&$ 0.4746 \pm 0.0069 $ & $ 0.1308 \pm 0.0037 $ \\
		\hline
		\multirow{2}{*}{\texttt{AGE}} 
		&{{64}}   & $ 0.5205 \pm 0.0312 $ & $ 0.2225 \pm 0.026 $& $ 0.5041 \pm 0.02 $ & $ 0.1955 \pm 0.0155 $&$ 0.4618 \pm 0.0196 $ & $ 0.164 \pm 0.0283 $\\
		&{{128}}   & $ 0.517 \pm 0.0334 $ & $ 0.2133 \pm 0.0208 $& $ 0.5107 \pm 0.0183 $ & $ 0.2003 \pm 0.0066 $& $ 0.4654 \pm 0.0262 $ & $ 0.1663 \pm 0.0335 $ \\
		\hline
        \multirow{2}{*}{\texttt{DANE}} 
		&{{64}}   & $ 0.6136 \pm 0.0356 $ & $ 0.2854 \pm 0.0143 $& $ 0.5503 \pm 0.025 $ & $ 0.2324 \pm 0.0161 $& $ 0.4903 \pm 0.0462 $ & $ 0.1776 \pm 0.04 $\\
		&{{128}}   & $ 0.7295 \pm 0.0457 $ & $ 0.4309 \pm 0.0378 $& $ 0.7064 \pm 0.0219 $ & $ 0.4089 \pm 0.026 $& $ 0.6287 \pm 0.0292 $ & $ 0.3181 \pm 0.0314 $\\
		\hline
		\bottomrule
	\end{tabular}}
\end{table*}
It is clear from the tables that the proposed \texttt{GAGE} significantly outperforms the baselines in both micro and macro F1 score, where \texttt{T-PINE} usually comes second. In the \texttt{Wikipedia} dataset there are instances where \texttt{T-PINE} is slightly better in micro F1 but \texttt{GAGE} is better in macro F1. Taking into consideration that the \texttt{Wikipedia} dataset consists of 19 classes and some classes are skewed, macro F1 score is far more significant in this dataset. Note that \texttt{Graph-AE} and \texttt{Graph-VAE} show in general very weak classification performance and especially for the $F=256$ in \texttt{BlogCatalog} they fail to produce acceptable results. We also notice that some baselines, that take attributes into account, produce weaker results compared to \texttt{Deepwalk} that only uses connectivity information. This is due to the fact that the considered datasets have missing attributes and certain baselines failed to be efficient under this challenging setting.
\vspace{-0.4cm}
\subsection{Link prediction}
The proposed embeddings are also tested for link prediction -- see supplementary material. We observed that \texttt{GAGE} achieves high prediction performance, but is sometimes outperformed by graph encoders and auto-encoders as \cite{kipf2016semi,gao2018deep,cui2020adaptive}. The reason is that graph encoders and auto-encoders treat the unobserved links as unknown rather than non-existing. This benefits link prediction but on the downside renders these approaches task-specific and results in weak classification performance. On the contrary, \texttt{GAGE} is a global approach, offers elite performance in both tasks and overall produces more informative node embeddings.
\begin{table*}
	\caption{Average score and standard deviation over 10 shuffles for BlogCatalog}
	\label{tab:Results BlogCatalog}
	\scalebox{0.8}{
	\begin{tabular}{ccccccccc}
		\toprule
		Algorithm & dimension  & micro (0.9) & macro (0.9)& micro (0.5) & macro (0.5)& micro (0.1) & macro (0.1)\\
		\midrule
		\multirow{2}{*}{\texttt{GAGE}} 
		&{{128}}   & $ 0.9233 \pm 0.009 $ & $ 0.9208 \pm 0.0095 $& $\mathbf{ 0.9191 \pm 0.0028} $ & $ \mathbf{0.9171 \pm 0.0027} $& $ \mathbf{0.8858 \pm 0.0101 }$ & $ \mathbf{0.8842 \pm 0.0098} $\\
		&{{256}}   & $ \mathbf{0.9538 \pm 0.0082} $ & $ \mathbf{0.9527 \pm 0.0082} $& $ \mathbf{0.9457 \pm 0.0017} $ & $ \mathbf{0.9447 \pm 0.0018} $& $ \mathbf{0.912 \pm 0.0066} $ & $ \mathbf{0.9109 \pm 0.0066}$\\
		\hline
		\multirow{2}{*}{ \texttt{T-PINE}} 
		&{{128}}   & $ \mathbf{0.9281 \pm 0.0087} $ & $ \mathbf{0.9263 \pm 0.0093} $&$ 0.9145 \pm 0.0045 $ & $ 0.913 \pm 0.0047 $ &$ 0.8577 \pm 0.0055 $ & $ 0.8563 \pm 0.0053 $ \\
		&{{256}}   &$ 0.9213 \pm 0.0098 $ & $ 0.9196 \pm 0.0097 $ &$ 0.9076 \pm 0.0043 $ & $ 0.9061 \pm 0.0044 $ & $ 0.8681 \pm 0.0048 $ & $ 0.867 \pm 0.0048 $\\
		\hline
		\multirow{2}{*}{\texttt{Deepwalk}} 
		&{{128}}   & $ 0.6937 \pm 0.0212 $ & $ 0.6802 \pm 0.0218 $
& $ 0.681 \pm 0.0056 $ & $ 0.673 \pm 0.0059 $& $ 0.6187 \pm 0.0083 $ & $ 0.6117 \pm 0.0081 $\\
&{{256}}   & $ 0.6923 \pm 0.0197 $ & $ 0.6796 \pm 0.0207 $& $ 0.6823 \pm 0.0051 $ & $ 0.6743 \pm 0.0054 $& $ 0.619 \pm 0.0089 $ & $ 0.6121 \pm 0.0086 $\\
		\hline
		\multirow{2}{*}{\texttt{Graph-AE}} 
		&{{128}}   & $ 0.2521 \pm 0.0128 $ & $ 0.179 \pm 0.0077 $& $ 0.2455 \pm 0.0086 $ & $ 0.1806 \pm 0.0133 $& $ 0.2547 \pm 0.0107 $ & $ 0.1454 \pm 0.0154 $\\
		&{{256}}   & $-$ & $-$& $-$& $-$& $-$& $-$\\
		\hline
		\multirow{2}{*}{\texttt{Graph-VAE}} 
		&{{128}}   & $ 0.5306 \pm 0.01 $ & $ 0.4896 \pm 0.0119 $& $ 0.5182 \pm 0.0092 $ & $ 0.4754 \pm 0.0128 $& $ 0.467 \pm 0.0149 $ & $ 0.4204 \pm 0.021 $\\
		&{{256}}   & $-$ & $-$& $-$& $-$& $-$& $-$\\
		\hline
		\multirow{2}{*}{\texttt{TADW}} 
		&{{128}}   & $0.8504 \pm0.0106$& $0.8483\pm0.012$& $0.8464\pm 0.0043$& $0.8442\pm0.0044$&$0.8296 \pm0.0046$& $0.8284\pm0.0042$\\
		&{{256}}   & $ 0.8485 \pm 0.0102 $ & $ 0.8466 \pm 0.0118 $& $ 0.8446 \pm 0.0041 $ & $ 0.8424 \pm 0.0042 $&$ 0.829 \pm 0.0048 $ & $ 0.8278 \pm 0.0042 $\\
		\hline
		\multirow{2}{*}{\texttt{DGI}} 
		&{{128}}   & $ 0.6017 \pm 0.0136 $ & $ 0.5624 \pm 0.0136 $& $ 0.5786 \pm 0.0149 $ & $ 0.527 \pm 0.0208 $&$ 0.3693 \pm 0.0583 $ & $ 0.2761 \pm 0.0602 $\\
		&{{256}}   & $ 0.6163 \pm 0.0175 $ & $ 0.5642 \pm 0.0188 $& $ 0.595 \pm 0.0157 $ & $ 0.5376 \pm 0.0179 $& $ 0.3236 \pm 0.0729 $ & $ 0.2235 \pm 0.0683 $\\
		\hline
		\multirow{2}{*}{\texttt{AGE}} 
		&{{128}}   & $ 0.7279 \pm 0.0174 $ & $ 0.7219 \pm 0.0183 $& $ 0.7109 \pm 0.0063 $ & $ 0.7053 \pm 0.0067 $& $ 0.7279 \pm 0.0174 $ & $ 0.7219 \pm 0.0183 $\\
		&{{256}}   & $ 0.726 \pm 0.0147 $ & $ 0.7195 \pm 0.0154 $& $ 0.7115 \pm 0.0052 $ & $ 0.7057 \pm 0.0055 $& $ 0.6708 \pm 0.0097 $ & $ 0.658 \pm 0.0105 $ \\
		\hline
		\multirow{2}{*}{\texttt{DANE}} 
		&{{128}}   & $ 0.4821 \pm 0.0126 $ & $ 0.4637 \pm 0.0153 $& $ 0.4654 \pm 0.0118 $ & $ 0.4364 \pm 0.0126 $& $ 0.4821 \pm 0.0126 $ & $ 0.4637 \pm 0.0153 $\\
		&{{256}}   & $ 0.6748 \pm 0.0149 $ & $ 0.6597 \pm 0.016 $& $ 0.6488 \pm 0.01 $ & $ 0.6207 \pm 0.015 $& $ 0.4206 \pm 0.0546 $ & $ 0.3407 \pm 0.0715 $ \\
		\hline
		\bottomrule
	\end{tabular}}
\end{table*}
\vspace{-0.3cm}
\subsection{Sensitivity analysis and running time}
We examined the effect of parameter $\lambda$ in the performance of \texttt{GAGE} for node classification and link prediction. The details are relegated to the supplementary material due to space limitations. In a nutshell, we observe that classification performance is consistent for $\lambda\in[0.1,0.9]$ and the best performance is usually achieved for $\lambda\in[0.5,0.9]$. Regarding link prediction, the best results are achieved when $\lambda=1$ and the performance deteriorates as lambda decreases.

We also measured the running time required for our proposed \texttt{GAGE} and the baselines to produce $128$-dimensional embeddings for the three datasets. The results are presented in Table \ref{tab:time}. \texttt{DANE} requires additional time (indicated after the plus sign) to compute random walks. It is clear that the proposed \texttt{GAGE} is the fastest for \texttt{WebKB} and \texttt{BlogCatalog}, whereas \texttt{TADW} is the fastest for \texttt{Wikipedia}.
\begin{table}
  \caption{Running time (sec)}
  \label{tab:time}
  \scalebox{0.67}{
  \begin{tabular}{cccccccccc}
    \toprule
  Dataset& \texttt{GAGE}&\texttt{T-PINE}&\texttt{Deepwalk}&\texttt{G-AE} &\texttt{G-VAE} & \texttt{TADW} & \texttt{DGI}& \texttt{AGE} & \texttt{DANE}\\
    \midrule
    \texttt{Wiki} & 32.2& 79.5 & 267 & 47.9 & 49.7 & \textbf{8.1}&42.7& 202.1 &155+1597.2\\
    \hline
    \texttt{WebKB} & \textbf{2.2}& 80.1 & 73.8 & 20 & 20 & 2.7&10.7&14.2&51.9+464.1\\
    \hline
    \texttt{BCatalog} & \textbf{17.7}& 628 & 653.8 & 347.6 & 340.2 & 63.4&269&1458.1&416.1+4111.8\\
    \hline
  \bottomrule
\end{tabular}}
\end{table}
\vspace{-0.2cm}
\section{Conclusions}
In this paper we proposed \texttt{GAGE}, a novel tensor-based approach for unsupervised node embedding of attributed networks. \texttt{GAGE} leverages the favorable properties of multi dimensional scaling and canonical polyadic decomposition and provides embeddings that preserve the geometry of both network connectivity and attributes. Although the proposed approach works with distance matrices rather than the original adjacency and attributes the algorithm can still exploit the sparsity structure of the graph and the attributes and admits a scalable and lightweight implementation. Experiments with real world benchmark networks showcase the effectiveness of the proposed \texttt{GAGE} on downstream tasks.
\appendix
\section{Appendix: Efficient CPD computations for MDS tensor}
We are given an adjacency matrix $\bm Y_1 = \bm S_{\mathcal{G}}\in\{0,1\}^{N\times N}$ and a matrix of node attributes $\bm Y_2 = \mathcal{\bm A}\in\mathbb{R}^{N\times d}$. We are interested in computing the CPD of tensor $\underline{\bm X}$ with:
\vspace{-0.1cm}
\begin{align}
\underline{\bm X}(:,:,1)=\bm X_1 = \left(\bm I-\frac{1}{N}\bm 1 \bm 1^T\right) \bm Y_1\bm Y_1^T\left(\bm I-\frac{1}{N}\bm 1 \bm 1^T\right),\\    
\underline{\bm X}(:,:,2)=\bm X_2 = \left(\bm I-\frac{1}{N}\bm 1 \bm 1^T\right) \bm Y_2\bm Y_2^T\left(\bm I-\frac{1}{N}\bm 1 \bm 1^T\right)
\end{align}
The objective of this Appendix is to show how to perform this CPD computation by exploiting the special sparsity structure and without instantiating a dense $\underline{\bm X}$.
\subsection{\texttt{GAGE-EVD}}\label{app:GAGE-evd}
The first step of \texttt{GAGE} algorithm involves an eigenvalue decomposition. The bottleneck operation is:
\vspace{-0.2cm}
\begin{equation}
    \bm V\bm\Sigma\bm V^T\leftarrow\text{EVD}\left(\bm{X}^{(1)^T}\bm{{X}^{(1)}},F\right)
    \vspace{-0.2cm}
\end{equation}
First let us observe the structure of matrix $\bm{X}^{(1)^T}\bm{{X}^{(1)}}$. Note that $\bm{{X}^{(1)}}=\begin{bmatrix}
\bm X_1\\ \bm X_2\end{bmatrix}$, and $\bm X_1,~\bm X_2$ are both symmetric matrices.
\vspace{-0.1cm}
\begin{align}
    &\bm{X}^{(1)^T}\bm{{X}^{(1)}}= \begin{bmatrix}
\bm X_1^T \bm X_2^T\end{bmatrix} \begin{bmatrix}
\bm X_1\\ \bm X_2\end{bmatrix} = \bm X_1^T\bm X_1 + \bm X_2^T\bm X_2=\\
&\left(\bm I-\frac{1}{N}\bm 1 \bm 1^T\right) \bm Y_1\bm Y_1^T\left(\bm I-\frac{1}{N}\bm 1 \bm 1^T\right) \bm Y_1\bm Y_1^T\left(\bm I-\frac{1}{N}\bm 1 \bm 1^T\right)+\\&\left(\bm I-\frac{1}{N}\bm 1 \bm 1^T\right) \bm Y_2\bm Y_2^T\left(\bm I-\frac{1}{N}\bm 1 \bm 1^T\right)\bm Y_2\bm Y_2^T\left(\bm I-\frac{1}{N}\bm 1 \bm 1^T\right),
\end{align}
\vspace{-0.1cm}
since $\left(\bm I-\frac{1}{N}\bm 1 \bm 1^T\right)\left(\bm I-\frac{1}{N}\bm 1 \bm 1^T\right)=\left(\bm I-\frac{1}{N}\bm 1 \bm 1^T\right)$. To compute the EVD of $\bm{X}^{(1)^T}\bm{{X}^{(1)}}$ we resort to the orthogonal iterations method \cite{golub2013matrix}. The steps are summarized as follows:
\begin{itemize}
    \item Initialize $\bm Q_0\in\mathbb{R}^{N\times F}:$ orthogonal matrix\\
    repeat:
    \item $\bm W_k=\left(\bm I-\frac{1}{N}\bm 1 \bm 1^T\right) \bm Y_1\bm Y_1^T\left(\bm I-\frac{1}{N}\bm 1 \bm 1^T\right) \bm Y_1\bm Y_1^T\left(\bm I-\frac{1}{N}\bm 1 \bm 1^T\right)\bm Q_{k-1} + \left(\bm I-\frac{1}{N}\bm 1 \bm 1^T\right) \bm Y_2\bm Y_2^T\left(\bm I-\frac{1}{N}\bm 1 \bm 1^T\right)\bm Y_2\bm Y_2^T\left(\bm I-\frac{1}{N}\bm 1 \bm 1^T\right)\bm Q_{k-1}$
    \item $\bm Q_k \leftarrow$ QR $(\bm W_k)$\\
until convergence
\end{itemize}
It is clear that the above procedure does not instantiate $\bm X_1,~\bm X_2$ and works directly with $\bm Y_1,~\bm Y_2$. In the first step of the loop every computation is either a sparse or rank 1 multiplication which can be performed efficiently. The computationally more intensive computation lies in the QR computation of matrix $\bm W_k$. The complexity of this step is $\mathcal{O}(NF^2)$ which is linear in the number of nodes.

\vspace{-0.2cm}
\subsection{Sparsity aware \texttt{GAGE}}\label{app:GAGE}
Now we study the ALS updates in \texttt{GAGE} algorithm. The update for $\bm U$ can be written as:
\vspace{-0.2cm}
\begin{align}
\bm U \leftarrow \text{solve} \left(\left(\bm C^T\bm C)\ast(\bm U^{'^T}\bm U^{'}\right)\right)\bm U^T = \left(\bm C\odot\bm U^{'}\right)^T\bm X^{(1)}.
\vspace{-0.2cm}
\end{align}
The matrix-matrix multiplication in the right hand side exploits the special structure of $\bm X^{(1)}$:
\vspace{-0.2cm}
\begin{align}\label{eq:KR}
&\left(\bm C\odot\bm U^{'}\right)^T\bm X^{(1)} =\begin{bmatrix}
\bm U^{'} \text{diag}\left(\bm C(1,:)\right)\\
\bm U^{'} \text{diag}\left(\bm C(2,:)\right)\\
\end{bmatrix}^T\begin{bmatrix}
\bm X_1\\ \bm X_2
\end{bmatrix}=\nonumber\\&\sum_{k=1}^{2}\text{diag}\left(\bm C(k,:)\right)\bm U^{'^T}\left(\bm I-\frac{1}{N}\bm 1 \bm 1^T\right) \bm Y_k\bm Y_k^T\left(\bm I-\frac{1}{N}\bm 1 \bm 1^T\right).
\vspace{-0.2cm}
\end{align}
It follows that the number of flops required to compute \eqref{eq:KR} is $\mathcal{O}(sF)$, where $s=s_1+s_2$ and $s_1,~s_2$ are the number of non-zeros in $\bm Y_1, \bm Y_2$ respectively. Furthermore, $\bm X_1,~\bm X_2$ are not being instantiated. The same principles hold for the update of $\bm U^{'}$:   
\begin{align}
\bm U^{'} \leftarrow \text{solve} \left(\left(\bm C^T\bm C)\ast(\bm U^T\bm U\right)\right)\bm U^{'^T} = \left(\bm C\odot\bm U\right)^T\bm X^{(2)}.
\vspace{-0.2cm}
\end{align}
\begin{align}
&\left(\bm C\odot\bm U\right)^T\bm X^{(2)} = \begin{bmatrix}
\bm U \text{diag}\left(\bm C(1,:)\right)\\
\bm U \text{diag}\left(\bm C(2,:)\right)
\end{bmatrix}^T\begin{bmatrix}
\bm X_1\\ \bm X_2
\end{bmatrix}=\nonumber\\&\sum_{k=1}^{2}\text{diag}\left(\bm C(k,:)\right)\bm U^T\left(\bm I-\frac{1}{N}\bm 1 \bm 1^T\right) \bm Y_k\bm Y_k^T\left(\bm I-\frac{1}{N}\bm 1 \bm 1^T\right).
\end{align}
The update of $\bm C$ can be written as:
\vspace{-0.2cm}
\begin{align}
\bm C \leftarrow \text{solve} \left(\left(\bm U^{'^T}\bm U^{'})\ast(\bm U^T\bm U\right)\right)\bm C^T = \left(\bm U^{'}\odot\bm U\right)^T\bm X^{(3)}.
\vspace{-0.2cm}
\end{align}
To avoid instantiating $\bm U^{'}\odot\bm U$, we observe that:
\begin{align}\label{last}
\left(\bm U^{'}\odot\bm U\right)^T\bm X^{(3)} =
\begin{bmatrix}
\bm U(:,1)^T \bm X_1 U^{'}(:,1),\bm U(:,1)^T \bm X_2 U^{'}(:,1)\\
\vdots\\
\bm U(:,F)^T \bm X_1 U^{'}(:,F),\bm U(:,F)^T \bm X_2 U^{'}(:,F)\\
\end{bmatrix}
\end{align}
The operation in \eqref{last} avoids storing $\bm U^{'}\odot\bm U$. Furthermore, the formula in \eqref{eq:MDS} is used for $\bm X_1,~\bm X_2$, which exploits the structure of $\bm Y_1,~\bm Y_2$ and does not instantiate $\bm X_1,~\bm X_2$. The overall operation can be computed efficiently in $\mathcal{O}(max\{NF,sF\})$ flops.



%

\newpage
\bibliographystyle{ACM-Reference-Format}
\bibliography{mybib.bib}

\newpage
\appendix

\section{Supplementary material}

\subsection{Link prediction}\label{app:link}
We test the performance of the competing algorithms in the link prediction task. To do that we remove 50$\%$ of the edges for each network and then run the embedding algorithms. We form a testing set of the removed edges along with an equal number of randomly sampled non-edges. Then we compute $\bm e_i^T\bm e_j$ for each $i,j$ edge in the testing set and rank the edges according to $\bm e_i^T\bm e_j$. Higher ranked edges are more likely to have a link. To assess the performance of the baselines we measure the area under ROC curve (AUC) and Average Precision (Avg. Prec.). The results are presented in Table \ref{tab:Results link} and are averaged over 5 shuffles. We observe that for \texttt{Wikipedia}, the proposed \texttt{GAGE} and the autoencoders work similarly and \texttt{AGE} is the best. In the \texttt{WebKB} network \texttt{DANE} works the best, whereas in \texttt{BlogCatalog} \texttt{Graph-VAE} and \texttt{Graph-AE} outperform \texttt{GAGE} and the baselines. However, taking into consideration that in node classification task \texttt{GAGE} works markedly better, we conclude that \texttt{GAGE} produces more informative node embeddings.
\begin{table*}
	\caption{Average score and standard deviation over 5 shuffles for link prediction}
	\label{tab:Results link}
	\scalebox{1}
	{\begin{tabular}{ccccccc}
		\toprule
		  & \multicolumn{6}{c}{Dataset}\\
		\cline{2-7}
		 Algorithm &  \multicolumn{2}{c}{Wikipedia}& \multicolumn{2}{c}{WebKB}& \multicolumn{2}{c}{BlogCatalog}\\
		 &   AUC& Avg. Prec.& AUC& Avg. Prec.& AUC& Avg. Prec.\\
		\midrule
		{ \texttt{GAGE}} &   $0.8405 \pm 0.0018$& $0.8819 \pm 0.0017$& ${0.8604} \pm {0.0078}$& ${0.8347} \pm {0.0112}$&${0.7201} \pm {0.0013}$& ${0.7589} \pm {0.0077}$\\
		\hline
	{ \texttt{T-PINE}}  &   $0.8208 \pm 0.0069$& $0.8767 \pm 0.0044$&$0.7134 \pm 0.0109$& $0.7308 \pm 0.0121$&$0.6472 \pm 0.0056$& $0.6638 \pm 0.0034$\\
		\hline
	{\texttt{Deepwalk}} &   $0.7965 \pm 0.0056$& $0.8254 \pm 0.0044$&$0.6104 \pm 0.0145$& $0.6646 \pm 0.0128$&$0.6645 \pm 0.0049$& $0.6891 \pm 0.0064$\\
		\hline
	{\texttt{Graph-AE}} &   $0.8250\pm 0.0040$& $0.8833\pm 0.0043$& $0.8037\pm 0.0287$& $0.8335\pm 0.0225$&$\mathbf{0.8231} \pm 0.0222$& $0.8202 \pm 0.0367$\\
		\hline
	{\texttt{Graph-VAE}} &   ${0.8479}\pm {0.0073}$& ${0.8949}\pm {0.0047}$& ${0.8014}\pm {0.0375}$& ${0.8314}\pm {0.0284}$&${0.8218}\pm {0.0100}$& $\mathbf{0.8248}\pm {0.0164}$\\
		\hline
	\texttt{TADW}&   $0.7087 \pm 0.0028$& $0.7722 \pm 0.0032$&  $0.7966 \pm 0.0160$& $0.8178 \pm 0.0186$&$0.5351 \pm 0.0014$& $0.5317 \pm 0.0008$\\
		\hline
		\texttt{DGI}&   $0.8262 \pm 0.0020$& $0.8409 \pm 0.0019$&  $0.7778 \pm 0.0045$& $0.8179 \pm 0.0032$&$0.7434 \pm 0.0021$& $0.7404 \pm 0.0003$\\
		\hline
		\texttt{AGE}&$\mathbf{0.9173} \pm 0.0025$& $\mathbf{0.9151} \pm 0.0048$&  $0.9040 \pm 0.0037$& $0.8612 \pm 0.0098$&$0.7747 \pm 0.0102$& $0.7533 \pm 0.0097$\\
		\hline
		\texttt{DANE}&$0.8228 \pm 0.0032$& $0.8346 \pm 0.0025$&  $\mathbf{0.9201} \pm 0.0100$& $\mathbf{0.8715} \pm 0.0088$&$0.6347 \pm 0.0220$& $0.6526 \pm 0.0378$\\
		\hline
		\bottomrule
	\end{tabular}}
\end{table*}

\subsection{Sensitivity analysis}\label{app:sens}
In this subsection we examine the effect of parameter $\lambda$ in the performance of the proposed \texttt{GAGE} embeddings for node classification and link prediction.

\begin{figure}[ht]
	\centering
	\caption{Effect of $\lambda$ on Wikipedia node classification}
	\begin{subfigure}[b]{0.49\linewidth}
		\includegraphics[width=\textwidth]{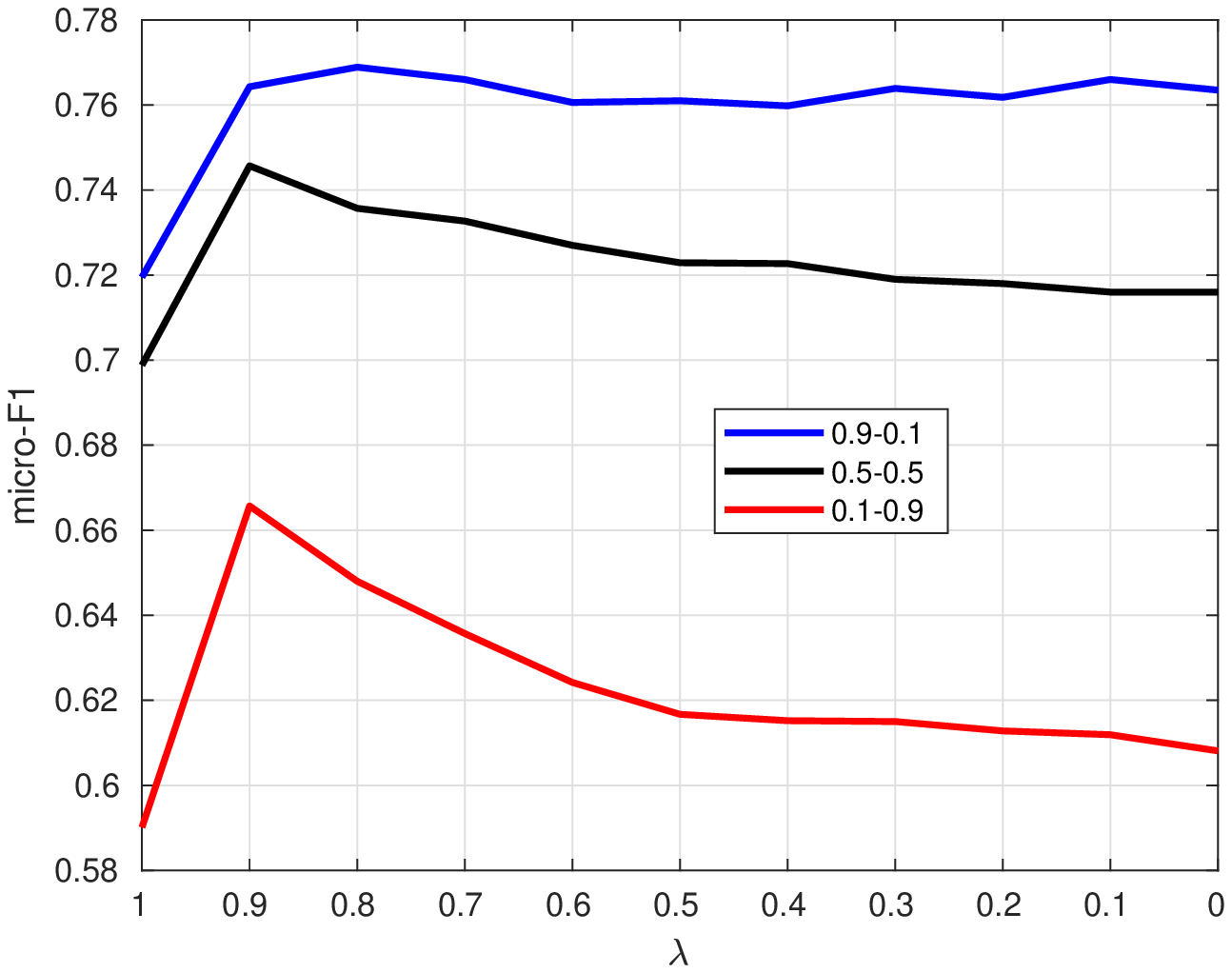}
		\caption{micro-F1 score}
	\end{subfigure}
	\begin{subfigure}[b]{0.49\linewidth}
		\includegraphics[width=\textwidth]{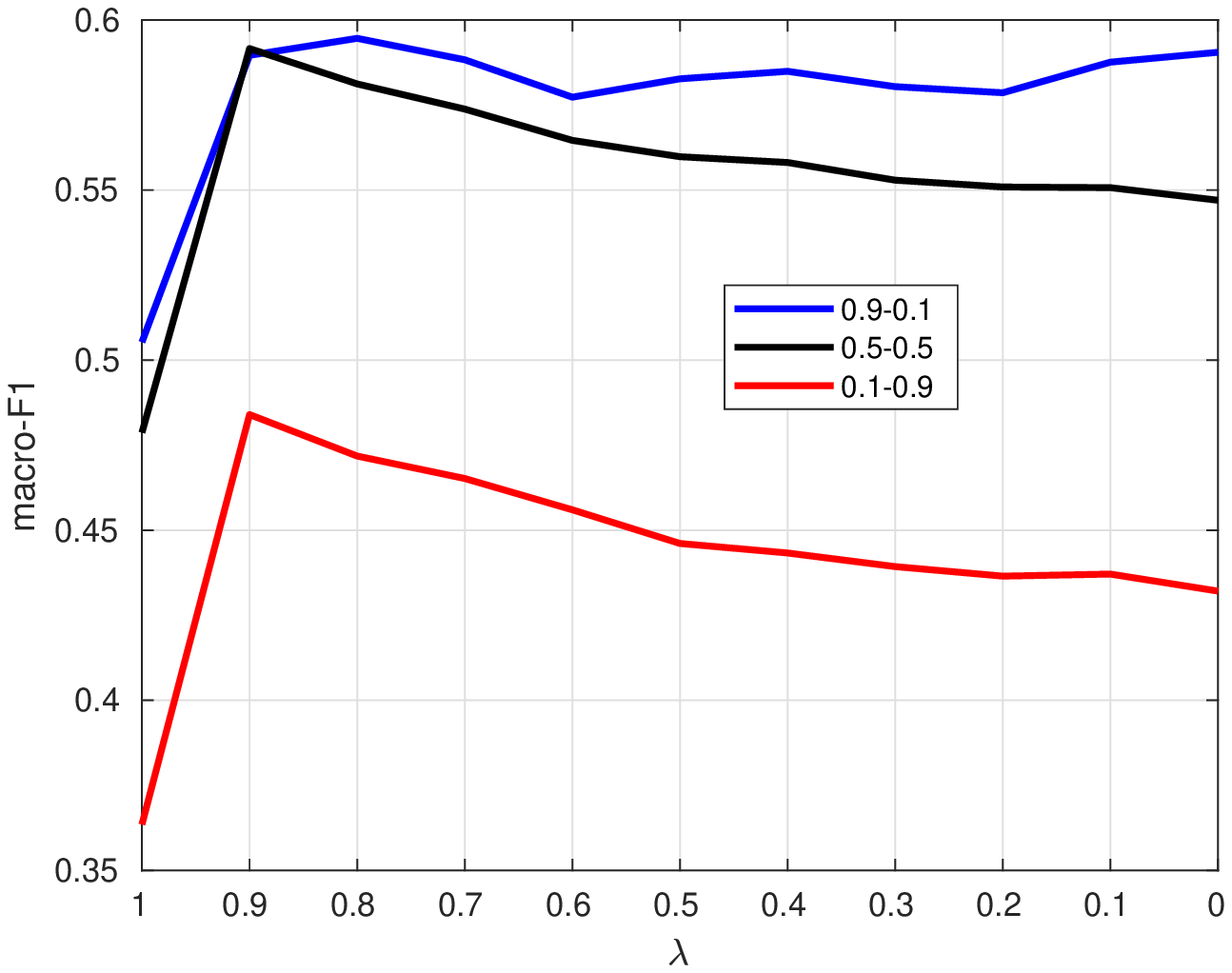}
		\caption{macro-F1 score}
	\end{subfigure}
	\label{fig:sens_NC_wiki}
\end{figure}

\begin{figure}[ht]
	\centering
	\caption{Effect of $\lambda$ on \texttt{WebKB} node classification}
	\begin{subfigure}[b]{0.49\linewidth}
		\includegraphics[width=\textwidth]{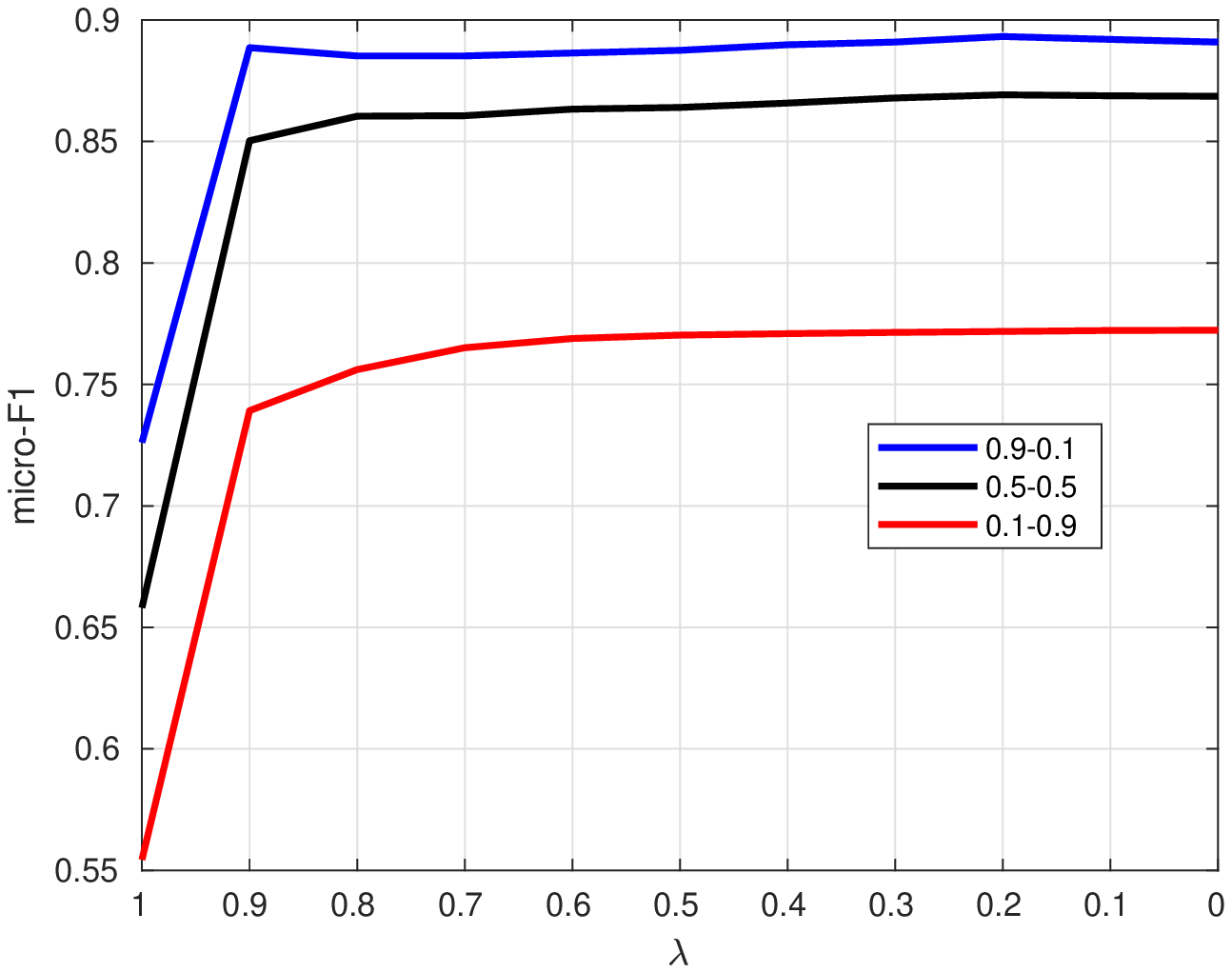}
		\caption{micro-F1 score}
	\end{subfigure}
	\begin{subfigure}[b]{0.49\linewidth}
		\includegraphics[width=\textwidth]{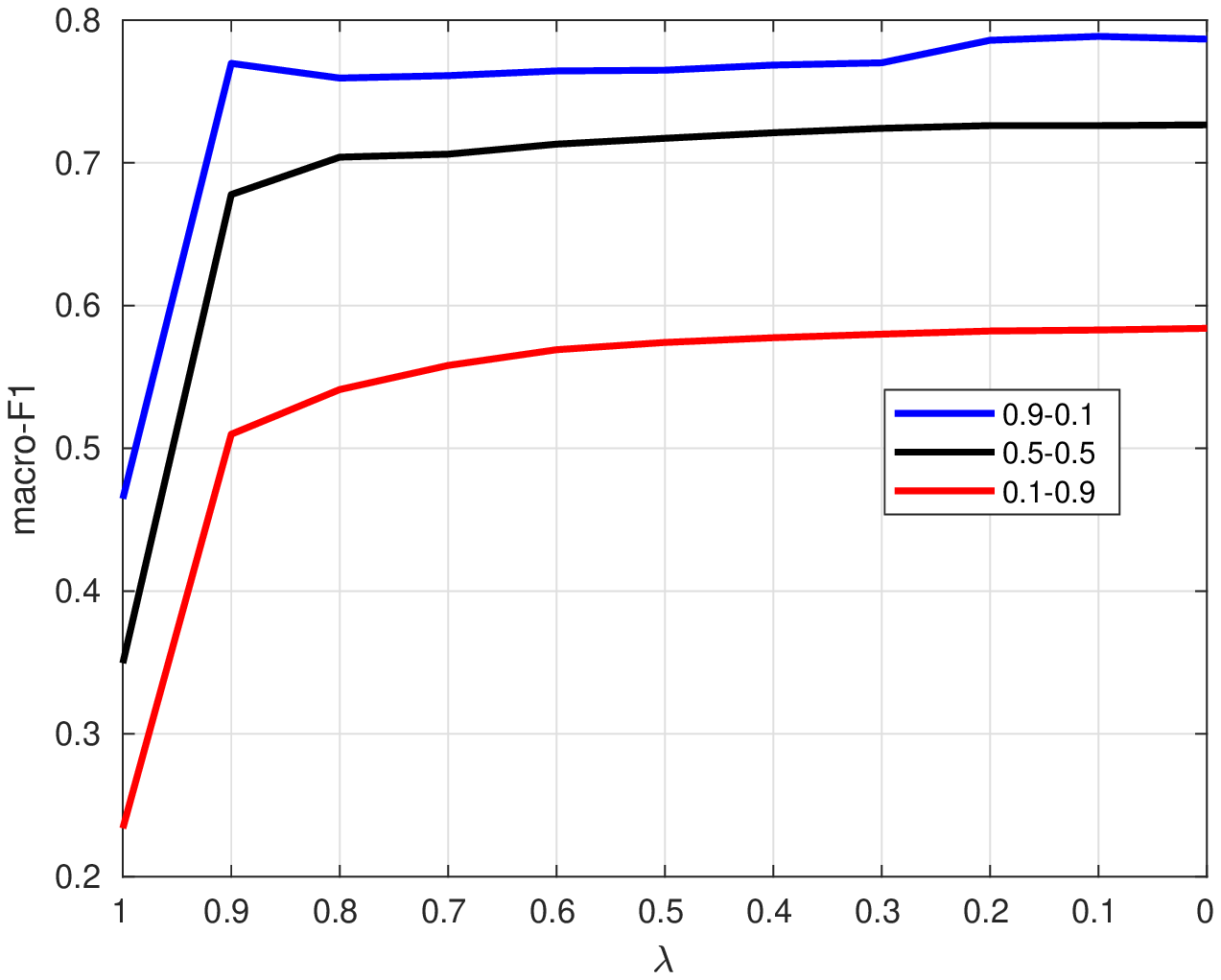}
		\caption{macro-F1 score}
	\end{subfigure}
	\label{fig:sens_NC_WB}
\end{figure}

\begin{figure}[ht]
	\centering
	\caption{Effect of $\lambda$ on \texttt{BlogCatalog} node classification}
	\begin{subfigure}[b]{0.49\linewidth}
		\includegraphics[width=\textwidth]{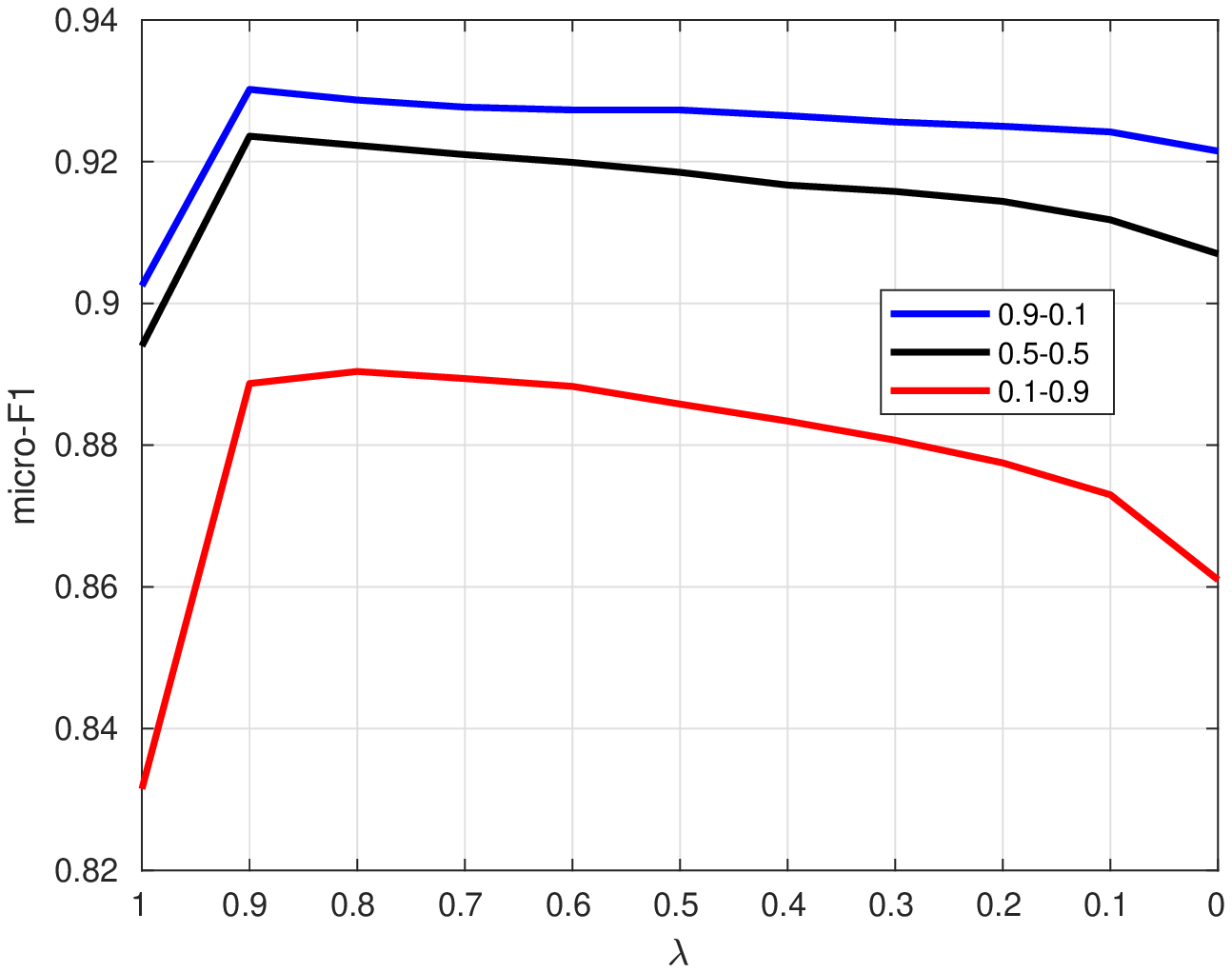}
		\caption{micro-F1 score}
	\end{subfigure}
	\begin{subfigure}[b]{0.49\linewidth}
		\includegraphics[width=\textwidth]{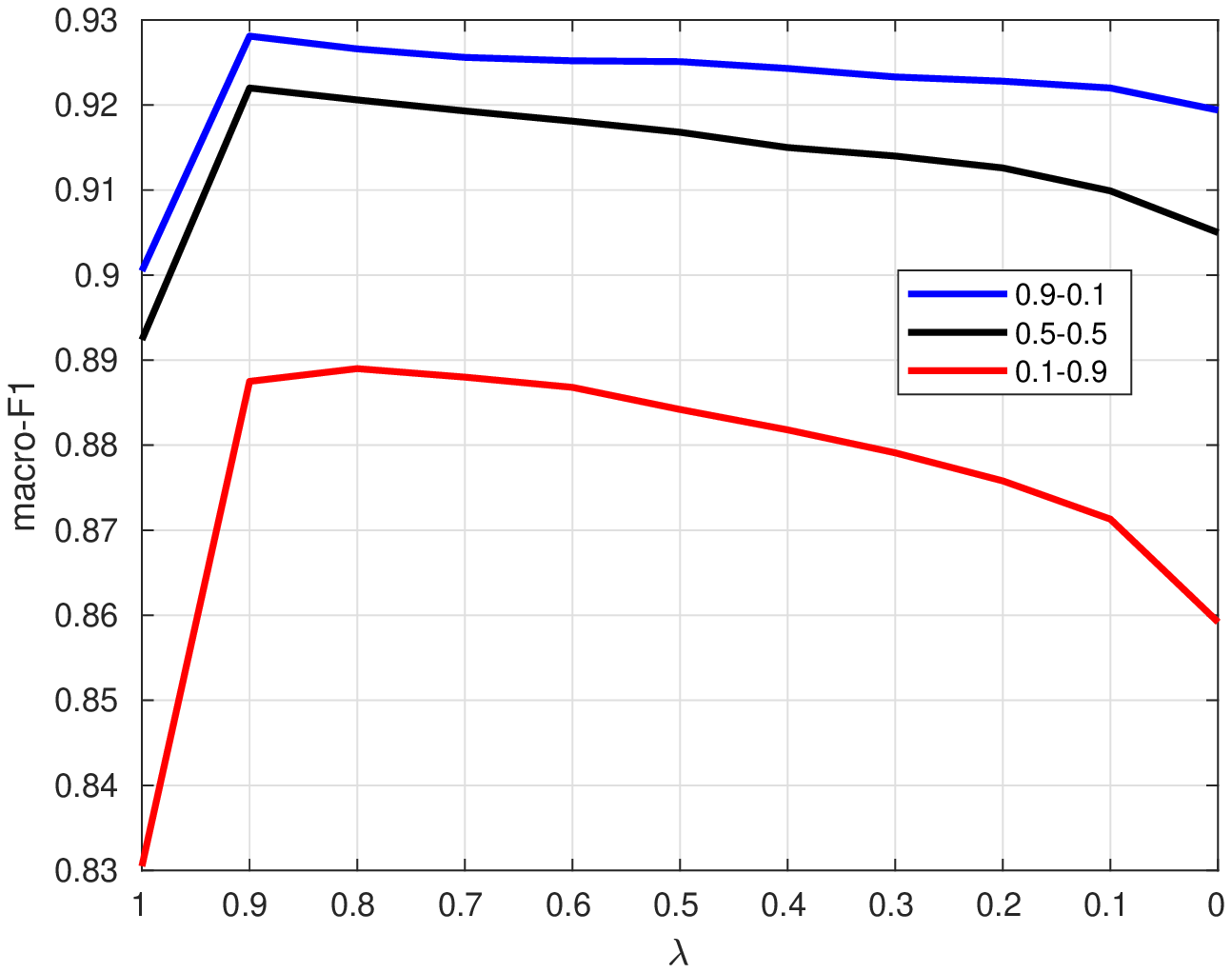}
		\caption{macro-F1 score}
	\end{subfigure}
	\label{fig:sens_NC_BC}
\end{figure}

First, we test the effect of $\lambda$ on node classification. We set the embedding dimension equal to $F=128$ and vary $\lambda$ from $1$ to $0$ with step equal to $0.1$. We measure micro-F1 and macro-F1 scores for $90-10,~50-50$ and $10-90$ training-testing splits. Recall that high values of $\lambda$ aim to preserve the network geometry associated with the connectivity information, whereas low values of $\lambda$ better preserve the attribute distances. The results for \texttt{Wikipedia}, \texttt{WebKB} and \texttt{BlogCatalog} are presented in Figs. \ref{fig:sens_NC_wiki}, \ref{fig:sens_NC_WB} and \ref{fig:sens_NC_BC} respectively. The classification performance is consistent for $\lambda\in[0.1,0.9]$ and the best performance is usually achieved for $\lambda\in[0.5,0.9]$. When $\lambda=1$ the focus is solely on the graph and  classification performance is weaker compared to all other values. This stresses the importance of network attributes in node representation learning and graph node classification.
\begin{figure}[ht]
	\centering
	\caption{Effect of $\lambda$ on link prediction}
	\begin{subfigure}[b]{0.49\linewidth}
		\includegraphics[width=\textwidth]{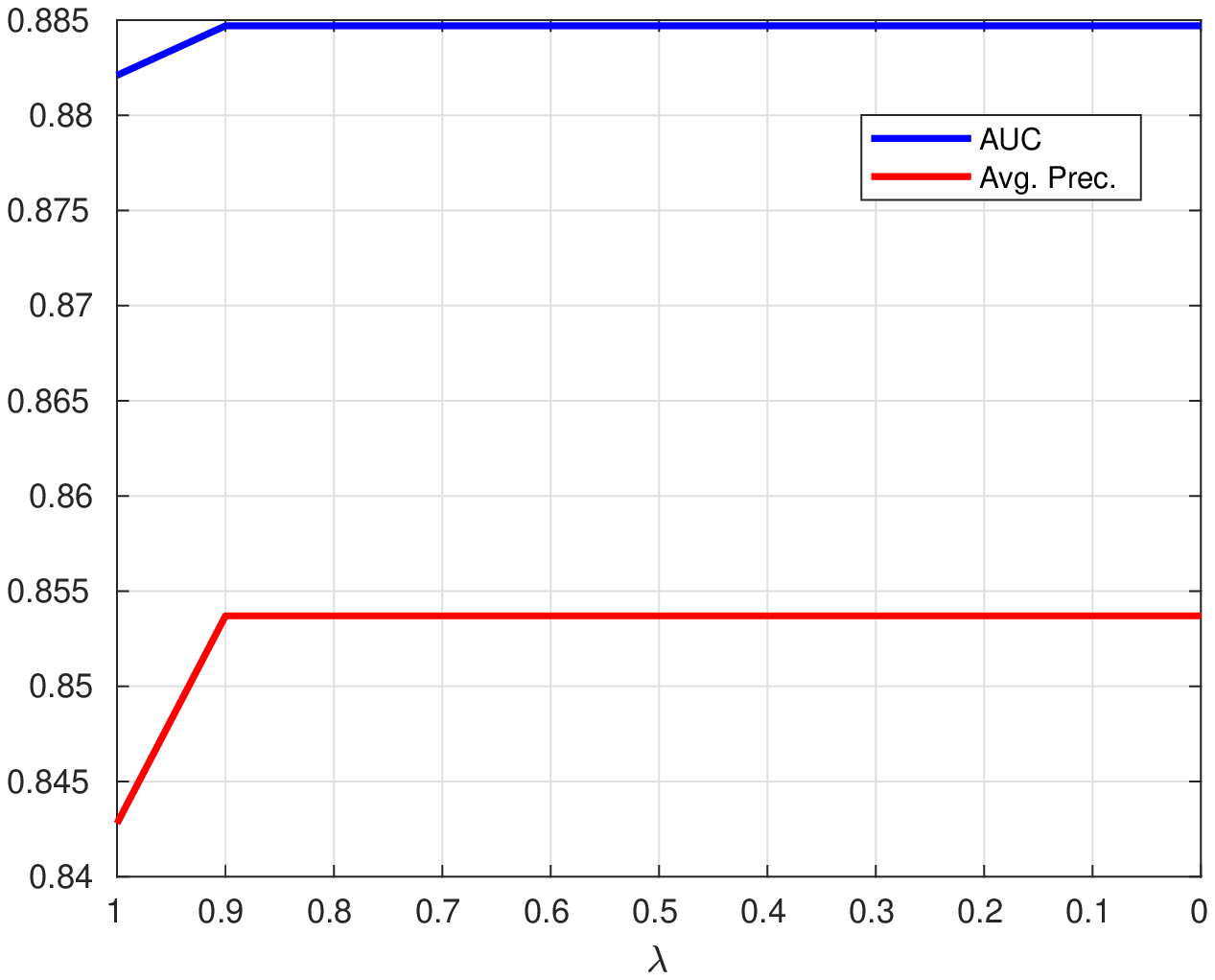}
		\caption{Wikipedia}
	\end{subfigure}
	\begin{subfigure}[b]{0.49\linewidth}
		\includegraphics[width=\textwidth]{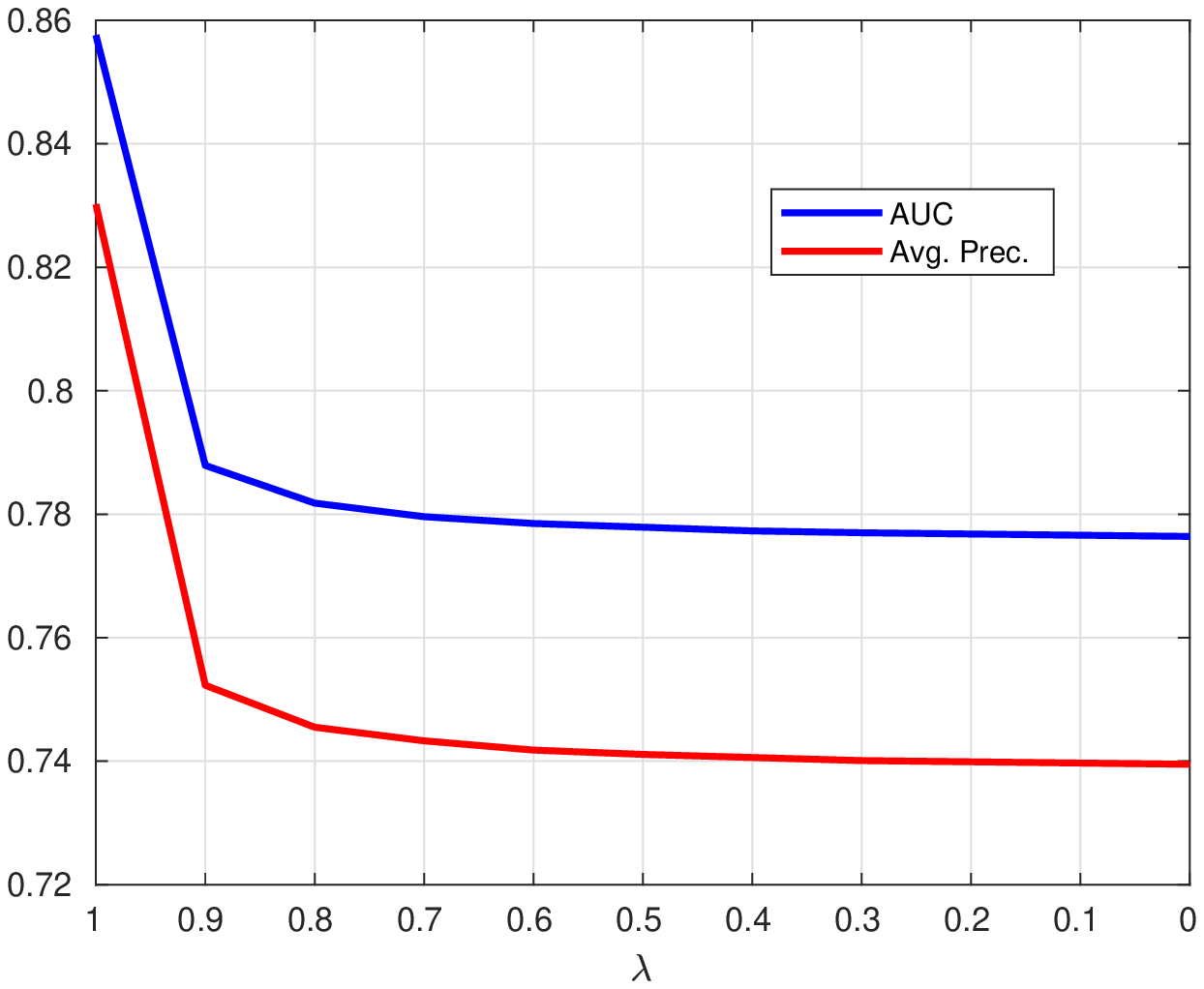}
		\caption{WebKB}
	\end{subfigure}
	\begin{subfigure}[b]{0.49\linewidth}
		\includegraphics[width=\textwidth]{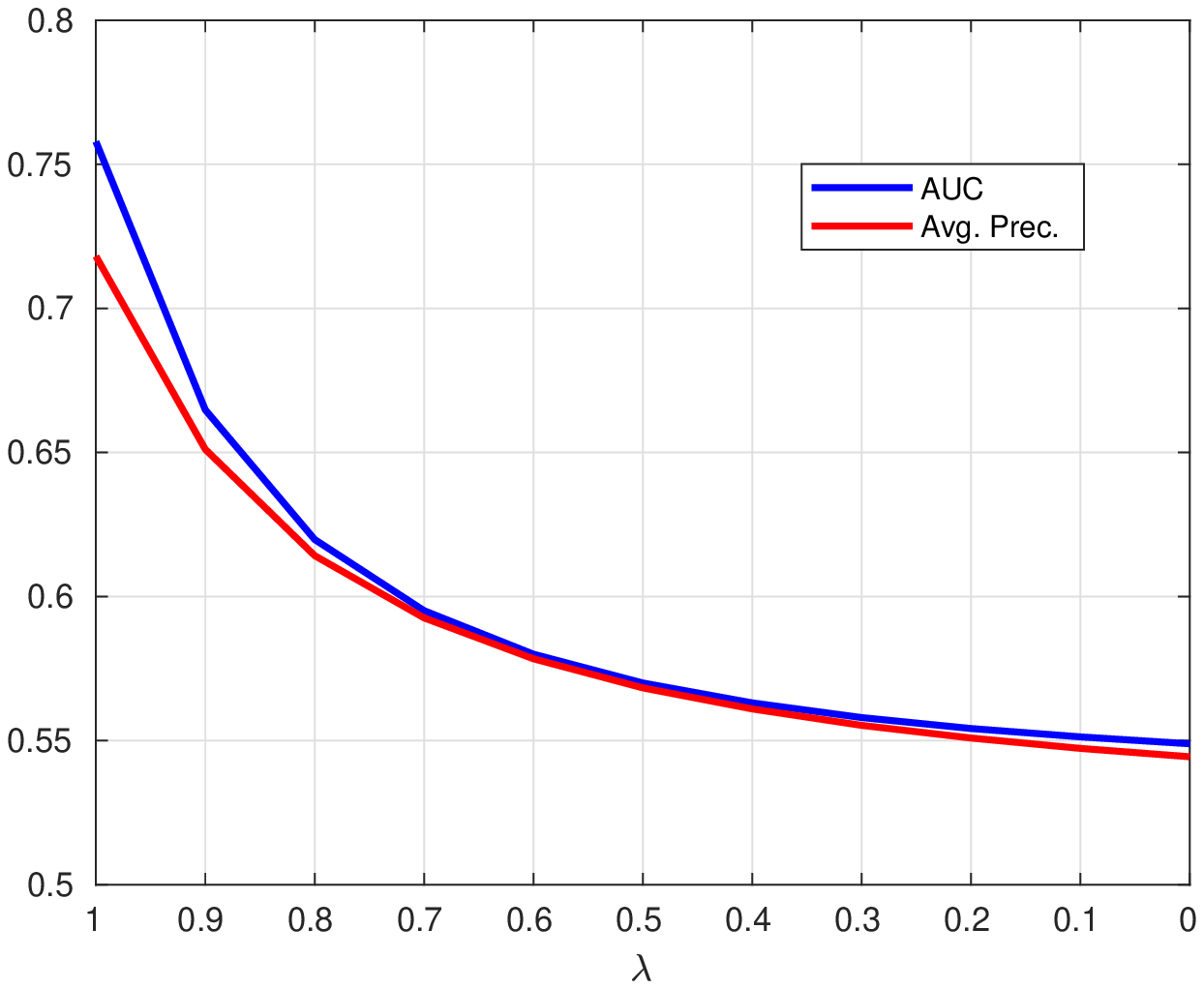}
		\caption{BlogCatalog}
	\end{subfigure}
	\label{fig:sens_LP}
\end{figure}

Next we examine the effect of parameter $\lambda$ in link prediction. In this direction we vary $\lambda$ from $1$ to $0$ with step equal to $0.1$, as before, and measure the AUC and Average Precision. The embedding dimension is set to $F=64,128,256$ for \texttt{WebKB}, \texttt{Wikipedia} and \texttt{BlogCatalog} respectively. The results are presented in Fig. \ref{fig:sens_LP}. For \texttt{BlogCatalog} the performance is consistent across all values of $\lambda$. For \texttt{Wikipedia} and \texttt{WebKB} we observe that better link prediction is achieved when $\lambda=1$ and the performance deteriorates as lambda decreases. This expected as potential links affect the graph geometry of the network and with $\lambda=1$ we focus on preserving the connectivity distances between the nodes.

\end{document}